\documentclass[final]{elsart}

\usepackage{amsmath}
\usepackage{amssymb,bm}
\usepackage{graphicx}
\usepackage{color}

\makeatletter
\newif\if@restonecol
\makeatother

\usepackage{algorithm2e}
\DeclareMathAlphabet\mathpzc{OT1}{pzc}{m}{it}
\let\mathcal=\mathpzc
\def\E{{\mathbb E}}
\def\P{{\mathbb P}}
\def\H{{\mathbb H}}

\let\trueiiint=\iiint
\def\iiint{\mathop{\textstyle\trueiiint}\limits}
\def\intinfty{\int\limits_{\!\!-\infty\,\,}^{\,\,\infty\!\!}\kern-0.0em}
\def\iintinfty{\mathop{\int\!\!\int}\limits_{\!\!-\infty\,\,}^{\,\,\infty\!\!}\kern-0.0em}
\def\iiintinfty{\mathop{\int\!\!\int\!\!\int}\limits_{\!\!-\infty\,\,}^{\,\,\infty\!\!}\kern-0.0em}

\let\<=\langle
\let\>=\rangle
\let\^=\hat
\let\-=\mathbf

\def\circ{\ifmmode\mathchar"220E\else$\mathchar"220E$\fi}
\def\@#1{{\cal #1}}

\def\COV{\mathrm{COV}}

\journal{Elsevier}

\begin{document}
\centerline{}
\begin{frontmatter}



\title{Gaussian process surrogates for failure detection: a Bayesian experimental design approach}


\author[authorlabel1]{Hongqiao Wang}
\ead{wanghongqiao@sjtu.edu.cn}

\address[authorlabel1]{Institute of Natural Sciences and Department of Mathematics, Shanghai Jiao Tong University,
Shanghai 200240, China.}

\author[authorlabel2]{Guang Lin}
\ead{guanglin@purdue.edu}
\address[authorlabel2]{Department of Mathematics  {and School of Mechanical Engineering},
Purdue University,
150 N. University Street,
West Lafayette, IN 47907-2067, USA.}

\author[authorlabel3]{Jinglai Li}
\ead{jinglaili@sjtu.edu.cn}
\address[authorlabel3]{Institute of Natural Sciences, Department of Mathematics, and MOE Key Laboratory of Scientific and Engineering Computing, Shanghai Jiao Tong University, Shanghai 200240, China. (Corresponding author)}


\medskip
\begin{center}
\end{center}

\begin{abstract}
An important task of uncertainty quantification is to identify  {the probability of} undesired events, in particular, system failures,
caused by various sources of uncertainties.
In this work we consider the construction of Gaussian  {process} surrogates for failure detection and failure probability estimation.
In particular, we consider the situation that the underlying computer models are extremely expensive,
and in this setting, determining the sampling points in the state space is of essential importance.
We formulate the problem as an optimal experimental design for Bayesian inferences of the limit state (i.e., the failure boundary) and propose
an efficient numerical scheme to solve the resulting optimization problem.
In particular, the proposed limit-state inference method is capable of determining multiple sampling points at a time, and thus it is well
suited for problems where multiple computer simulations can be performed in parallel.
The  {accuracy and} performance of the proposed method is demonstrated
by both academic and practical examples.

\end{abstract}

\begin{keyword}
Bayesian inference, experimental design, failure detection, Gaussian processes, Monte Carlo, response surfaces, uncertainty quantification
\end{keyword}
\end{frontmatter}

\section{Introduction}\label{s:intro}

Real-life engineering systems are unavoidably subject to various uncertainties
such as material properties, geometric parameters,
boundary conditions and applied loadings.
These uncertainties may cause undesired events, in particular, system failures or malfunctions, to occur.
Accurate identification of failure region and evaluation of
failure probability of a given system is an essential task in many fields of engineering such as risk
management, structural design,  {reliability-based} optimization, etc.

Conventionally the failure probability is often computed by constructing linear or quadratic expansions of the system model around
the so-called most probable point, known as the first/second-order reliability method~(FORM/SORM); see e.g., \cite{ditlevsen1996structural,SchuellerPK04}
and the references therein.
It is well known that FORM/SORM may fail for systems with  {nonlinearity} or multiple failure regions.
The Monte Carlo (MC) simulation, which estimates the failure probability by repeatedly simulating the underlying system, is another popular method for solving such problems. The MC method makes no approximation to the underlying computer models and thus can be applied to any systems.
On the other hand, the MC method is notorious for its slow convergence,
and thus can become prohibitively expensive
 when the underlying computer model is
computationally intensive and/or the system failures are rare  {and each sample} requires to a full-scale
numerical simulation of the system.
To reduce the computational effort,  one can construct an computationally inexpensive approximation to the true model, and then evaluate
the approximate model in the MC simulations.  Such approximate models are also known as response surfaces,
surrogates, metamodels,  {and emulators, etc. These methods} are referred to as
 the response surface (RS) methods~\cite{Faravelli89,BucherB90,RajashekharE93,GaytonBL03,GuptaM04,Pulch_2008} in this work.
 The response surface  can often provide a reliable estimate of the failure probability, at a much lower computational cost than direct MC simulations.

In this work we are focused on a specific kind of RS, the Gaussian processes (GP) surrogates, also known as kriging in many fields of applications.
The GP surrogates have been widely used in machine learning~\cite{williams2006gaussian}, geostatistics
~\cite{oliver1990kriging},  engineering optimizations~\cite{simpson2001kriging}, and most recently, uncertainty quantifications~\cite{bilionis2012multi,bilionis2013multi}. 
A number of  {GP-based} methods have been also been successfully implemented for failure probability estimation~\cite{li2012bayesian,balesdent2013kriging}.
In this work we consider the situation where the underlying computer models are extremely expensive and one can only afford a very limited number of simulations.
In this setting, choosing the sampling points (i.e. the parameter values with which the simulation is performed) in the state space is of essential importance.
Determining the sampling points for GP can be cast as to optimally design computer experiments and considerable efforts~\cite{Krause2008} have been devoted to it.
Those methods aim to construct a surrogate that can accurately approximate the target function in the whole parameter space.
As will be explained later, in the failure probability estimation or failure detection problems, only the sign of the target function is used. 
Thus by requiring surrogates to be globally accurate, the methods may allocate considerable computational efforts to the regions not of interest,
and use much more model simulations than necessary.

Several methods have developed to determine sampling points for the failure probability estimation purposes.
Most of these methods consist of sequentially finding the ``best point'' as a result of a heuristic balance between predicted closeness to the limit state, and high prediction uncertainty, e.g. \cite{echard2011ak,bichon2008efficient}.
Such methods are shown to be effective in many applications, while
a major limitation is their point-wise  {sequential} nature, which  makes it  unsuitable for problems in which multiple computer simulations can be performed parallelly.
An exception is the stepwise uncertainty reduction (SUR) method developed in \cite{bect2012sequential,chevalier2014fast}, in which the authors proposed an optimal experimental design framework which determines multiple sampling points by minimizing the average variance of the failure probability.
It should be noted that the design criteria in the SUR method is particularly developed for the goal of estimating the failure probability only.
In practice, one is often not only interested in estimating the failure probability, but also identifying the events that can cause failures;
the latter demands a design criteria for the goal of detecting the limit state, i.e., the boundaries of the failure domain.
In this work, we recast the surrogate construction as a Bayesian inference to identify the limit state,
and based on that, we formulate an \emph{information-theoretic} {optimal experimental design}, which uses the relative entropy from the prior to the posterior as the design criteria,  to determine the sampling points.
We also present an efficient numerical scheme for solving the resulting optimal design problem, modified from the simulation-based method developed in
\cite{Huan2013}.
We compare the performance of the proposed limit-state inference (LSI) method with that of the SUR
 by numerical examples.


We note that another line of research in failure probability estimation is  {to} develop more efficient sampling schemes, such as the subset simulations~\cite{AuB01,AuB99}, importance sampling~\cite{melchers1989importance,Engelund93},
sequential Monte Carlo~\cite{cerou2012sequential}, the cross-entropy method~\cite{RubinsteinK_04,wang2015cross}, etc.
For practical engineering systems, computer simulations can be
extremely time consuming. In many cases, one can only afford very
limited number of simulations --- nothing beyond a few hundreds. In this
case, even the most effective sampling method is not applicable.
To this end, surrogates are needed even in those advanced sampling schemes and in particular the LSI method can be easily integrated into the aforementioned sampling schemes,
resulting in more efficient estimation schemes. Examples of combining surrogates and efficient sampling schemes include
  \cite{li2011efficient,li2012bayesian,balesdent2013kriging,dubourg2013metamodel}.

The rest of this paper is organized  {as following}. We first review
the preliminaries of our work in Section
\ref{s:setup}, including the mathematical formulation of failure probability computation and the GP surrogates. Our Bayesian experimental design framework and its numerical implementations are presented in Section~\ref{s:method}. 
Numerical examples
are presented in Section \ref{s:results} to demonstrate the
effectiveness of the proposed method, and finally  {Section~\ref{s:conclusion}} offers some closing remarks.

\section{Problem formulation} \label{s:setup}
\subsection{Failure detection and failure probability estimation}
Here we describe the failure probability estimation problem in a general setting.
We consider a probabilistic model where  $\-x$ is a $d$-dimensional random variable that represents the uncertainty in model and
the system failure is often defined using a real-valued function  $g(\cdot): R^d\rightarrow R$, which is known as the \emph{limit state function} or the
\emph{performance function}.
Specifically, the event of  failure is defined as  $g(\-x)<0$ and
as a result the failure probability is
\[P=\P(g(\-x)<0) = \int_{ \{\-x\in R^d | g(\-x)<0\}} I_g(\-x) p(\-x) d\-x=\int_{\-x\in R^d} I_g(\-x) p(\-x) d\-x,\]
where  $I_g(\-x)$ is an indicator function:
\[
I_g(\-x) = \left\{ \begin{array}{ll}
         1 &\quad \mbox{if $g(\-x) < 0$},\\
         0 & \quad \mbox{if $g(\-x) \geq 0$}; \end{array} \right.
\]
and $p(\-x)$ is the probability density function (PDF) of $\-x $.
In what follows we shall omit the integration domain when it is simply $R^d$.
This is a general definition for failure probability, which is used widely in many disciplines involving reliability analysis and risk management.
$P$ can be computed with the standard
 Monte Carlo estimation:
\begin{equation}
  \hat{P}= \frac{1}{n}\sum^n_{i=1}{I}_{g}(\-x_{i}),
\end{equation}
where samples $\-x_1,...,\-x_n$ are drawn from distribution $p(\-x)$.

In practice, many engineering systems require high reliability, namely the failure probability $P\ll 1$.
In this case, MC requires a rather large number of samples to produce a reliable estimate of the failure probability.
For example, for $P\approx 10^{-3}$, MC simulation requires $10^5$ samples to obtain an estimate with $10\%$ coefficient of variation.
On the other hand, in almost all practical cases, the limit state function $g(\-x)$ does not admit analytical expression and has to be
evaluated through expensive computer simulations, which renders the crucial MC estimation of the failure probability prohibitive.
To reduce the number of full-scale computer simulations, one can construct
a computationally inexpensive surrogate $G(\-x)$ to replace
the real function $g(\-x)$ in the MC estimation.
In this work we choose the Gaussian Process surrogates
and we provide a description of GP in the next section.

\subsection{Gaussian process surrogates}

The GP method constructs the approximation of $g(\-x)$ in a nonparameteric Bayesian regression framework~\cite{OHagan1978,williams2006gaussian}.
Specifically the target function $g(\-x)$ is cast as
\begin{equation}
g(\-x) = \mu(\-x)+\epsilon(\-x)
\end{equation}
 where $\mu(\-x)$ is a real-valued function and $\epsilon(\-x)$ is a  {zero-mean} Gaussian process whose covariance is specified by
a kernel $K(\-x,\-x')$, namely,
\[ \COV[\epsilon(\-x),\epsilon(\-x')] = K(\-x,\-x'). \]
The kernel $K(\-x,\-x')$ is positive semidefinite and bounded. Suppose that $N$ computer simulations of the function $g(\-x)$ are performed
 at parameter values $\-X^* := \left[\-x^*_1, \ldots \-x^*_n\right]$, yielding  function evaluations $\-y^* := \left[ {y}^*_1, \ldots {y}^*_n\right]$,
where \[{y}^*_i = g(\-x_i^*)\quad \mathrm{for} \quad i=1,\ldots,n.\]
Suppose we want to predict the function values at points $\-D := \left[\-x_1, \ldots \-x_{m}\right]$, i.e., $\-y=[y_1,\ldots y_m]$ where
$ {y_i=g(\-x_i)}$.
The posterior distribution of $\-y$ is
\begin{subequations}
\label{e:gp}
\begin{equation}
  \-y ~|~\-D,\-X^*, \-y^* \sim\mathcal{N}(\-u, \text{\sf C}), 
	\label{e:post}
\end{equation}
where the posterior mean $\-u = (u_1,\ldots u_m)$ is a $m$-dimensional vector with each element to be
\begin{equation}
u_j =  \mu(\-x_j) + \-r_j^T \text{\sf R}^{-1}(\-y^*-\bm\mu)  \quad\text{for} \quad {j=1...m}
\label{eq:postMean},
\end{equation}
where $\bm\mu = (\mu_1,\ldots,\mu_n)$ and  $\mu_i = \mu(\-x^*_i)$ for $i=1,\ldots,n$.
The posterior covariance matrix $\text{\sf C}$ is given by
\begin{equation}
\label{eq:postcovsimple}
(\text{\sf C}\,)_{j,j'} = K(\-x_j,\-x_{j'}) - \-r_j^T \, \text{\sf R}^{-1} \, \-r_{j'}.
\end{equation}
\end{subequations}
In the equations above, $\-r_j $ represents a $n$-dimensional vector whose $i$-th component is $K(\-x^*_i, \-x_j)$ and the matrix ${\text{\sf R}}$ is given by
$({\text{\sf R}})_{i,{i'}} = K(\-x^*_i, \-x^*_{i'})$ for $i,i'=1,\ldots, n$.

\section{The experimental design framework} \label{s:method}
A key question in constructing a GP surrogate is to determine the locations $\-D$ where the true limit state function $g(\-x)$ is evaluated,
which is often known as a design of computer experiments.
In this section, we show that determining the sampling locations can be translated
into a Bayesian experimental design problem whose goal is to find the
locations of limit state with a given prior distribution.

\subsection{Bayesian inference experiments}
We formulate a Bayesian inference problem based on the following argument.
In failure probability estimation, the limit state function $g$ is only used in the indicator
function $I_g(\-x)$ and so one is really interested in the sign of $g(\-x)$ rather than the precise value of it.
To this end, the essential task in constructing surrogates for the failure probability estimation is to learn about the boundary of the failure domain.
Let $\-z$ represents the boundary of the failure domain, i.e., the solutions of $g(\-x) = 0$.
We want to identify the boundary $\-z$ with Bayesian inference.
In this setting, the data is obtained by making observations, i.e., evaluate $g(\-x)$ at locations $\-D = (\-x_1,\-x_2,\ldots,\-x_n)$, resulting in data $\-y = (y_1,\ldots,y_n)$ with each $y_i = g(\-x_i)$.
The likelihood function $p(\-y |\-z,\-D)$ in this case is given by Eqs.~\eqref{e:gp} with $\-X^* = [\-z]$ and $\-y^*=[0]$.
We also need to choose a prior distribution for $\-z$, and without additional information, it is reasonable to assume that the prior is simply $p(\-z)$.
In this setting the posterior distribution of $\-z$ with data $\-y$ can be computed with the Bayes' theorem:
\[ p(\-z|\-y, \-D) = \frac{p(\-y |\-z,\-D) p(\-z)}{p(\-y|\-D)},\]
where $p(\-y|\-D)$ is the evidence.

\subsection{Optimal design criteria}
As is mentioned earlier, determining the locations $\-D$ can be formulated as a design of the Bayesian inference experiments.
Following the decision-theoretic optimal experimental design framework, an objective for
experimental design can be generally formed:
\begin{multline}
  U(\-D)=\int\int u(\-D,\-y, \-z)p(\-z, \-y|\-D)d\-z
  d\-y\\=\int \int u(\-D, \-y, \-z)p( \-z| {\-y},\-D)p( {\-y}|\-D)d\-z
  d\-y,
\end{multline}
where $u(\-D, {\-y}, {{\-z}})$ is a utility
function, $U(\-D)$ is the expected utility. The utility function $u$ should
reflect the usefulness of an experiment at conditions $\-D$, i.e., we
will get a particular value outcome $\-y$ at condition
$\-D$ by inputting a particular value of the parameters $\-z$. Since we do not know the exact value of $\-z$ and $\-y$ in advance, we take the expectation of $u$ over the joint
distribution of $\bf {{z}}$ and $\bf {y}$, resulting in the expected utility $U(\-D)$.
The optimal choice of $\-D$ then can be obtained by maximizing the expected utility of the design space $\Xi$:
\begin{equation}
  \-D^*=\arg\max\limits_{\-D\in \Xi}U(\-D). \label{e:maxU}
\end{equation}

A popular choice for the utility function is  {the} relative entropy, also known as the Kullback-Leibler divergence (KLD), between the prior and the posterior distributions.
For two distributions $p_A(\-z)$ and $p_B(\-z)$, the KLD from $p_A$ to $p_B$ is defined as
\begin{eqnarray}
  D_{KL}(p_A\parallel
  p_B)=\int p_A(\-z)\ln[\frac{p_A(\-z)}{p_B(\-z)}]d\-z=\mathbb{E}_A[\ln
\frac{p_A(\-z)}{p_B(\-z)}]
\end{eqnarray}
where  we define
$0\ln0\equiv 0$. This quantity is non-negative, non-symmetric, and
reflects the difference in information carried by the two
distributions.
When KLD is used in our inference problem, the utility function becomes
\begin{equation}
  u(\-D,\-y,\-z)\equiv
  \@D_{KL}(p(\cdot |\-y, \-D)\parallel p(\cdot))=\int
  p(\tilde{\-z}| {\-y},\-D)\ln \left[\frac {p(\tilde{\-z}| {\-y},\-D)}
  {p(\tilde{\-z})} \right]d\tilde{\-z}. \label{e:u-kld}
\end{equation}
The utility function $u(\-D,\-y,\-z)$ in Eq.~\eqref{e:u-kld} can be understood as the information gain
by performing experiments under conditions $\-D$, and larger value of $u$ implies that the experiment is more informative for parameter inference.
Note that the utility function in Eq.~\eqref{e:u-kld} is independent of $\-z$ and as such the expected utility $U(\-D)$ is reduced to
\begin{multline}
  U(\-D) =\int \@D_{KL}(p(\cdot |\-y, \-D)\parallel p(\cdot)) p(\-y |\-D) \, d\-y
 \\ =\int \int p(\-z| {\-y},\-D)\ln \left[\frac {p(\-z| {\-y},\-D)}
  {p(\-z)} \right]d\-z  \,p(\-y |\-D) \,d\-y, \label{e:Ud}
\end{multline}
where $\tilde{\bf {{z}}}$ in Eq.~\eqref{e:u-kld} is
replaced by $\-z$ for simplicity.
Next we discuss how to numerically solve the optimization problem Eq.~\eqref{e:maxU}, with $U(\-D)$ is given by Eq.~\eqref{e:Ud}.

\subsection{Numerical implementation}
Following the recommendation of \cite{Huan2013}, we use the simultaneous perturbation stochastic approximation (SPSA)
method,  to solve the optimization problem~\eqref{e:maxU}.
SPSA is a  {derivative-free} stochastic optimization method that was first proposed by Spall~\cite{spall1992multivariate,spall1998implementation},
and we provide the detailed algorithm of SPSA in Appendix A.
Note here that, since it is a  {derivative-free} method, the algorithm only uses the function value of $U(\-D)$.

Next we discuss the evaluation of $U(\-D)$.
To start, we re-write Eq.~\eqref{e:Ud} as
 \begin{align}
  U(\-D) &=\int \int p(\-z| {\-y},\-D)\ln \left[\frac {p(\-z| {\-y},\-D)}{p(\-z)} \right]d\-z  \,p(\-y |\-D) \,d\-y \nonumber
	 \\
	 & = -\frac12\int
  \ln |\text{\sf C}|
\,  p(\-z)\,d\-z  - \int \int
  \ln[p(\-y|\-D)] p(\-y|\-D) \,d\-y +Z \nonumber
  \\&  = -\frac12\E[\ln|\text{\sf C}|] + \H[p(\-y|\-D)] +Z
	\label{e:Ud3}
\end{align}
where $\H(\cdot)$ is the notation for entropy and $Z$ is a constant independent of $\-D$.
The detailed derivation of Eq.~\eqref{e:Ud3} is given in Appendix B.
Note that Eq.~\eqref{e:Ud3} typically has no closed-form expression and has to be evaluated with MC simulations.
Draw $m$ pairs of samples $\{(\-y_1,\-z_1),\ldots,(\-y_m,\-z_m)\}$ from $p(\-y,\-z|\-D)$, and
the MC estimator of $\E[\ln |\text{\sf C}|]$ is
\begin{equation}
\hat{U}_1 = \frac1m\sum_{i=1}^m |\text{\sf C}(\-z_i,\-D)|.
\end{equation}
Recall that in \cite{Huan2013},  the entropy term $\H[p(\-y|\-D)]$ is computed using a nested MC.
For efficiency's sake, we use the resubstitution method developed in \cite{ahmad1976nonparametric} to estimate the entropy.
The basic idea of the method is rather straightforward:
given a set of samples $\{\-y_1,\ldots,\-y_m\}$ of $p(\-y|\-D)$, one first computes an estimator of the density $p(\-y|\-D)$, say $\hat{p}(\-y)$,
with certain density estimation approach, and then estimates the entropy with,
\begin{equation}
\hat{U}_2 = \frac1m\sum_{i=1}^m \hat{p}(\-y_i).
\end{equation}
Theoretical properties of the resubstitution method are analyzed in \cite{joe1989estimation,hall1993estimation} and other entropy estimation methods can be found in \cite{beirlant1997nonparametric}.
In the original work~\cite{ahmad1976nonparametric}, the distribution $\hat{p}$ is obtained with kernel density estimation, which can become very costly when the dimensionality of $\-y$ gets high,
and to address the issue, we use Gaussian mixture based density estimation method~\cite{mclachlan2004finite}.
Without loss of generality we can simply set the constant $Z=0$ and thus an estimate of $U$ is simply
\begin{equation}
\hat{U} = \hat{U}_1+\hat{U}_2 \label{e:uhat},
\end{equation}
which will be used to provide function values in the SPSA algorithm.
Finally we reinstate that the numerical implementation in this work differs from a more general implementation outlined in \cite{Huan2013} in the following two aspects:
first, thanks to the Gaussianity of $p(\-y|\-D)$ , we can analytically integrate  {the} variable $\-y$ in the first integral on the right hand side of Eq.~\eqref{e:Ud3};
second, we use Gaussian mixture density estimation method to approximate $p(\-y|\-D)$ to avoid the nested MC integration.
We note that there are other methods such as \cite{long2013fast} that can be used to approximate the nested integral.

\subsection{Sequential optimal design}

Until this point we have presented our method under the assumption
that the sampling points are determined all in advance of performing computer simulations,
which is often referred to as an open-loop design.
In many applications, a more practical strategy is to choose the sampling points in a sequential fashion: determine a set of sampling points, perform simulations, determine another set of points based on the previous results, and so forth.
A sequential (close-loop) design can be readily derived from the open-loop version.
For computational efficiency, we adopt the ``greedy'' approach for the sequential design, while noting that
this approach can only obtain sub-optimal solutions in general.
 Simply speaking, the greedy sequential design iterates as follows until a prescribed stopping criteria is met:
 \begin{enumerate}
\item determine $n$ sampling points with an open-loop design;
\item perform simulations at the chosen sampling points;
\item use the simulation results to update $p(\-z)$ and $p(\-y|\-z,\-D)$;
\item return to step~(1).
\end{enumerate}
A key ingredient in the greedy procedure is that the prior $p(\-z)$ and the likelihood $p(\-z |\-z,\-D)$ are updated based on the simulation results in each stage.
To be specific, the prior distribution at step $k+1$ is the posterior distribution of $\-z$ at step~$k$:
\[p_{k+1}(\-z) = p_k(\-z | \-y, \-D).\]
The update of the likelihood is a bit more complicated.
Namely, let $\-X_k$ represent all the sampling points at which the function $g(\cdot)$ has been evaluated in the first $k$ stages and $\-y_k$ be
the associated function values, and the likelihood function $p_{k+1}(\-y|\-z,\-D)$ is once again given by Eqs.~\eqref{e:gp} while
\[\-X^* = [\-X_k,\-z]\quad \mathrm{and}\quad \-y^* = [\-y_k, 0].\]
Note that we do not have analytical expression for $p_{k+1}(\-z)$ or equivalently $p_k(\-z|\-y,\-D)$; however, our implementation only requires
samples drawn from $p_{k+1}(\-z)$ which are available from the previous iteration.
Another issue that should be noted is that, every time a new observation $(\-z,0)$ is added to $\-X^*$,
the posterior covariance needs to be updated.
Recomputing the covariance matrix for each sample can be rather time consuming,
and so here we use  {the} covariance matrix update method which computes the new covariance matrix based on the previous one.
 The detailed update formulas are given in \cite{emery2009kriging,chevalier2014corrected} and we shall not repeat it here.
Finally we also need to specify a stopping criteria for the algorithm and following the previous works we stop the iterations
when the difference the estimated probabilities in two consecutive iterations is smaller than a threshold value.

\section{Numerical examples} \label{s:results}
\subsection{Branin function}
\begin{figure}
\centerline{\includegraphics[width=.6\textwidth]{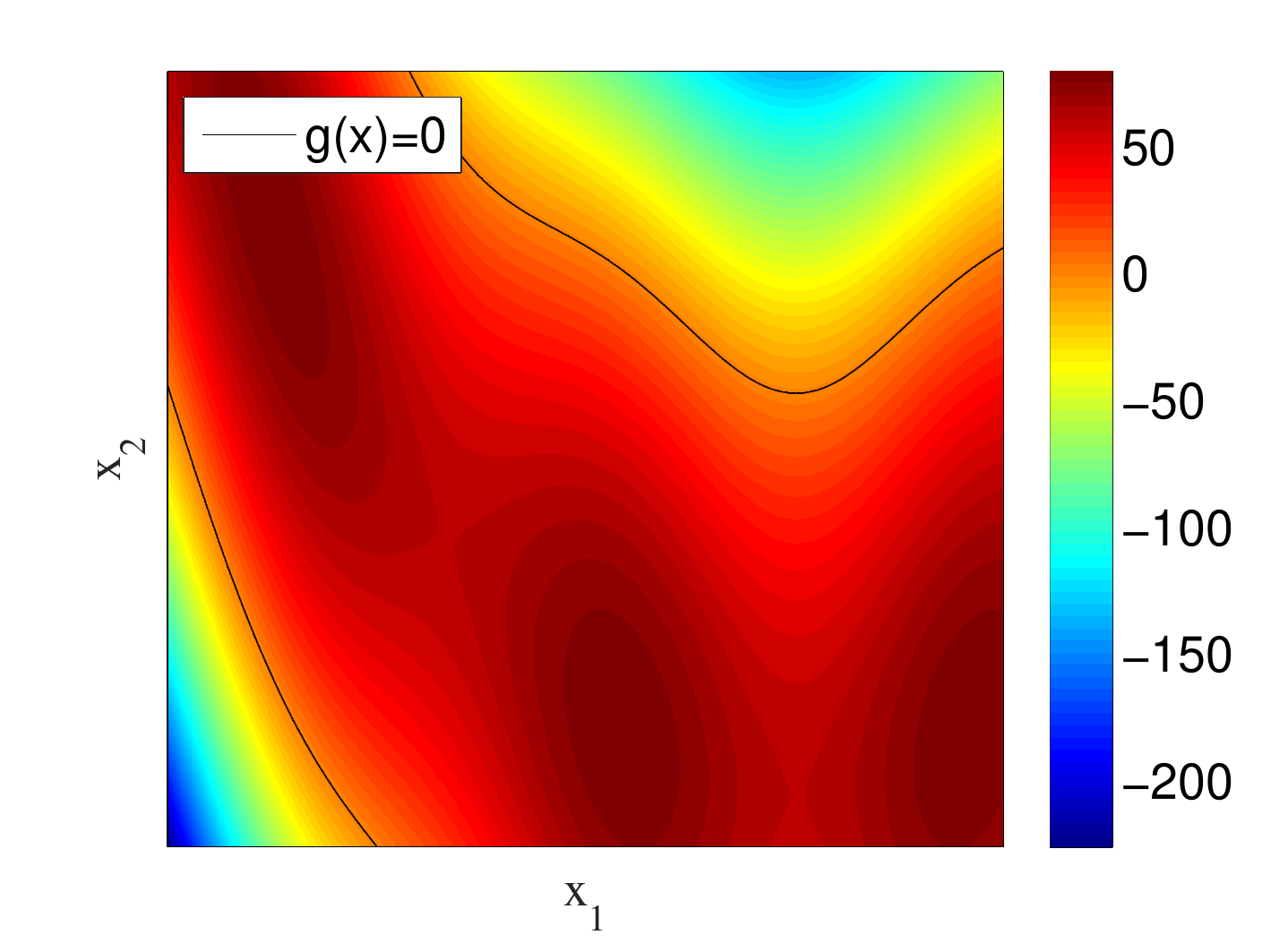}}
\caption{The rescaled Branin function.
The solid line is the limit state $g(\mathbf{x}) =0$.}\label{f:branin}
\end{figure}

To compare the performance of our method with SUR, we first test both methods on the rescaled Branin function
\begin{multline}
    g(x_1,x_2)=80-[(15x_2 - \frac5{4  \pi^2}  (15x_1-5)^2 + \frac5\pi (15x_1-5) - 6)^2 \\
		+ 10  (1 -\frac1{8 \pi})  \cos(15x_1-5)+10].
\end{multline}
which is plotted in Fig.~\ref{f:branin}.
Here we assume that $x_1$ and $x_2$ are independent and each  {follows} a uniform distribution in $[0,1]$.
The function is included in the R package KrigInv~\cite{Chevalier20141021} as an example for the SUR method
 and the numerical results suggest that SUR performs well for  {this numerical example}.

\begin{figure}
\centerline{\includegraphics[width=.45\textwidth]{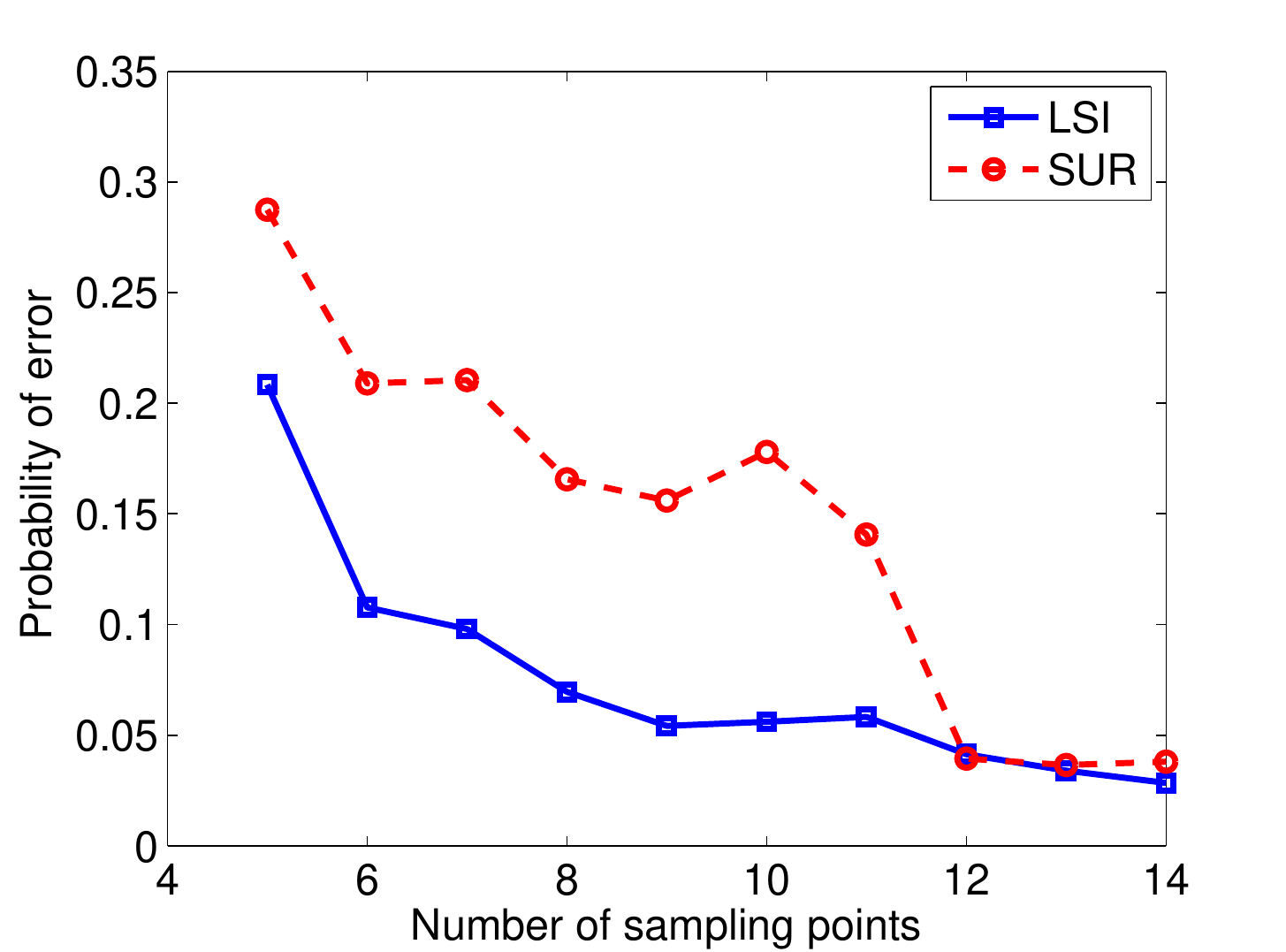}\includegraphics[width=.45\textwidth]{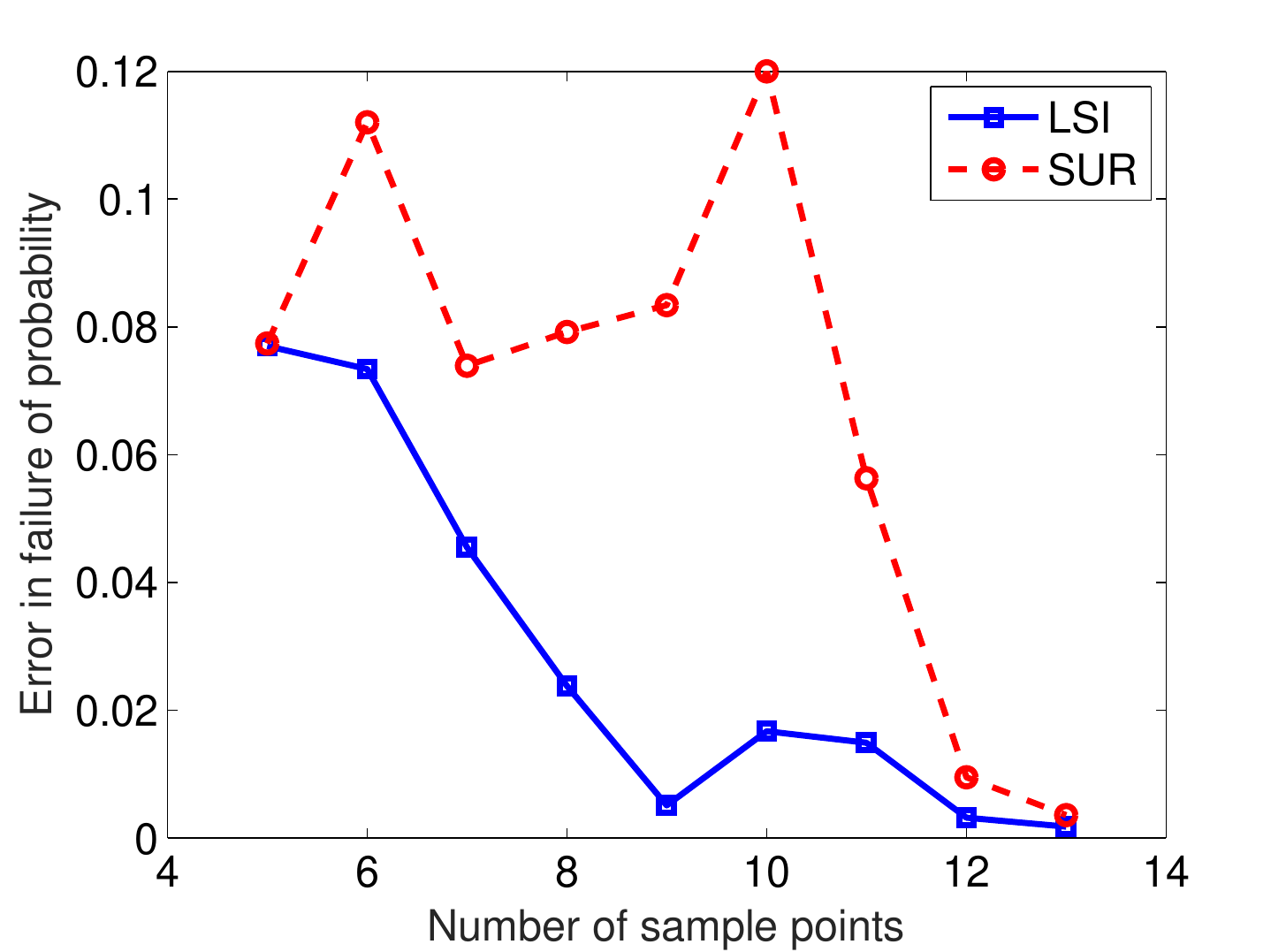}} 
\caption{Left: the probability of detection error plotted against the number of design points.
Right: the error in the failure probability estimates plotted against the number of design points.}\label{f:compare}
\end{figure}

We first perform a standard MC with 10,000 samples, resulting in a failure probability estimate of $0.256$, which will be used
as the ``true'' failure probability to validate our results.
In the experimental design, we first choose 4 points with the Latin hypercube method as the initial design points for both methods.
We then sequentially choose 9 points with both methods, where one point is determined in each iteration.
In the numerical experiment, we choose to use a squared exponential covariance:
\begin{equation}
K(\-x,\-x') = \alpha\exp( - \frac{\|\-x-\-x'\|_2^2}{\beta}),\label{e:secov}
\end{equation}
where we choose $\alpha=0.1$ and $\beta=1$ in this example.
If desired, the parameter values in Eq.~\eqref{e:secov} can also be obtained with maximum marginal likelihood method.
The prior mean is determined with a quadratic regression.
In all the three examples,  the algorithm is terminated after 50 iterations in the SPSA method, and in each iteration,
5000 samples are used to evaluate the value of $U(\-D)$.
We  draw 10000 samples from the uniform distribution, and each time a new design point is added, we use the resulting GP surrogates to detect
which points are in the failure region and which are not.
Using the detection results we compute the failure probability.
To compare the performance of the two methods, we compute two quantities as the indicators of performance. The first is the error between
the failure probability estimate by each method and the true failure probability which is estimated by standard MC.
The second is the probability that a point is mis-identified:
\[\P[\-x\in \{\-x| g(\-x)<0, \hat{g}(\-x)>0\}\cup \{\-x| g(\-x)>0, \hat{g}(\-x)<0\}], \]
where $\hat{g}(\-x)$ represents the surrogate.
In Fig.~\ref{f:compare}, we plot both indicators as a function of the number of design points for our method and the SUR.
The results of both indicators suggest that our LSI method seems to have a better performance than SUR in this example.

\subsection{Four branch system}
\begin{figure}
\centerline{\includegraphics[width=.6\textwidth]{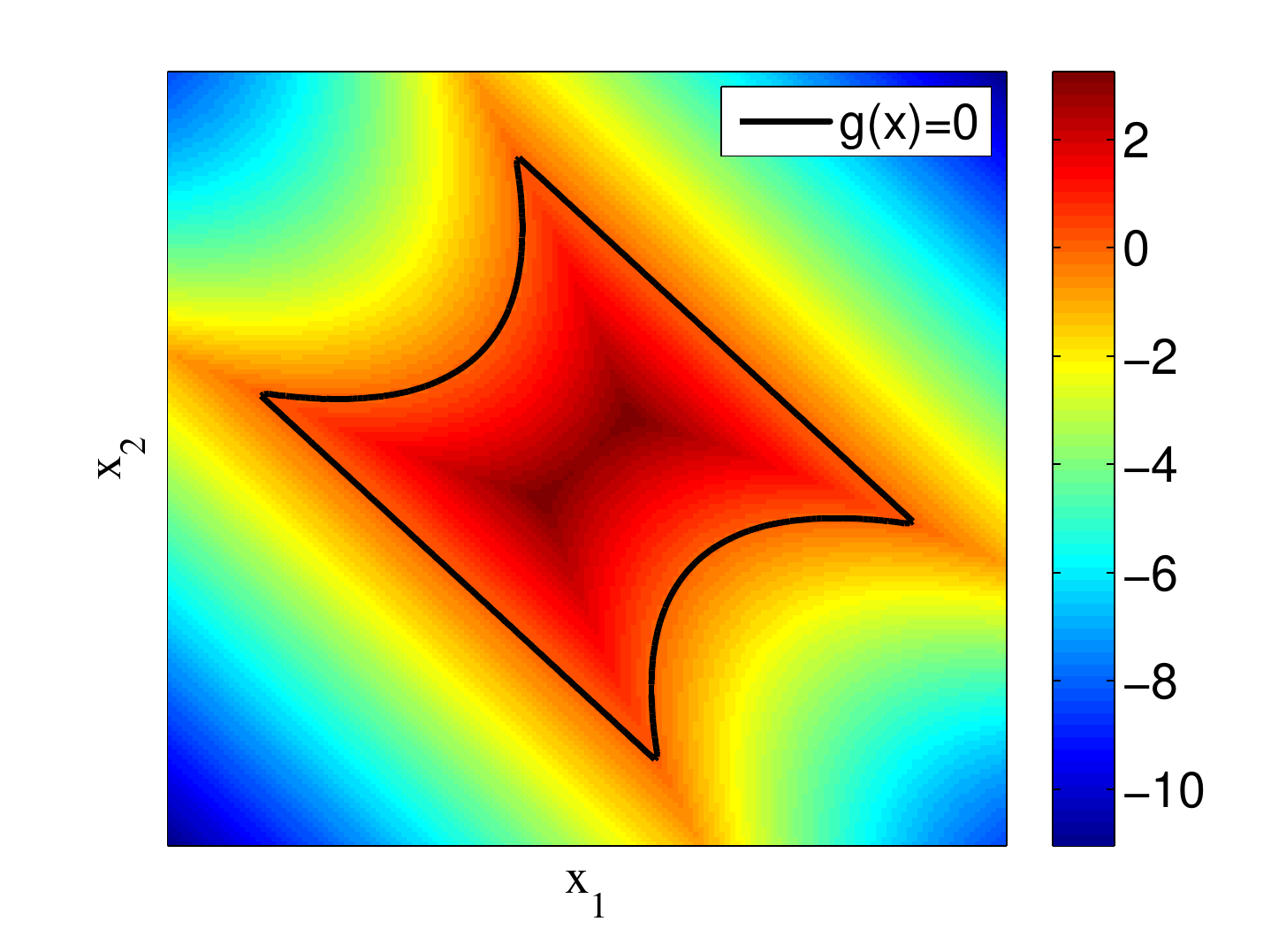}}
\caption{The limit state function of the four-branch model.
The solid lines in both figures are the limit state $g(\mathbf{x}) =0$.}\label{f:gfun}
\end{figure}

Our second example is the so-called four branch system, which is a popular benchmark test case in reliability analysis.
In this example the limit state function reads
\[
g(x_1,x_2) = \min\left\{
\begin{array}{l}
  3+0.1(x_1-x_2)^2-(x_1+x_2)/\sqrt{2}\\
  3+0.1(x_1-x_2)^2+(x_1+x_2)/\sqrt{2}\\
	(x_1-x_2)+7/\sqrt{2}\\
	(x_2-x_1)-7/\sqrt{2}
\end{array} \right\},
\]
which is shown in Fig.~\ref{f:gfun}.
The input random variable $x_1$ and $x_2$ are assumed to be independent and follow standard normal distribution.
We first compute the failure probability with a standard MC estimation of $10^{5}$ samples,
resulting an estimate of $2.34\times10^{-3}$, i.e., $234$ samples fall in the failure region.

\begin{figure}
\centerline{\includegraphics[width=.45\textwidth]{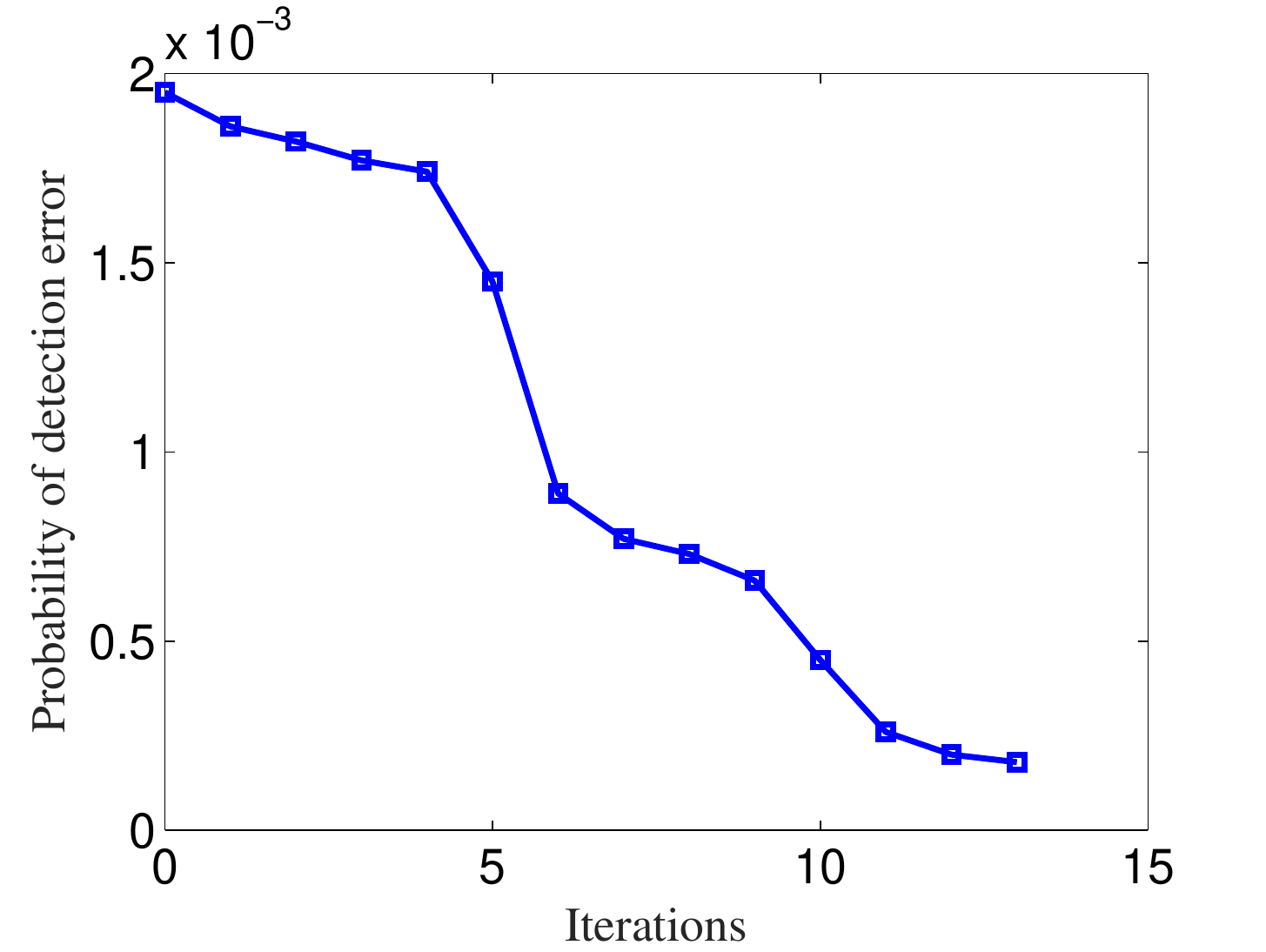}\includegraphics[width=.45\textwidth]{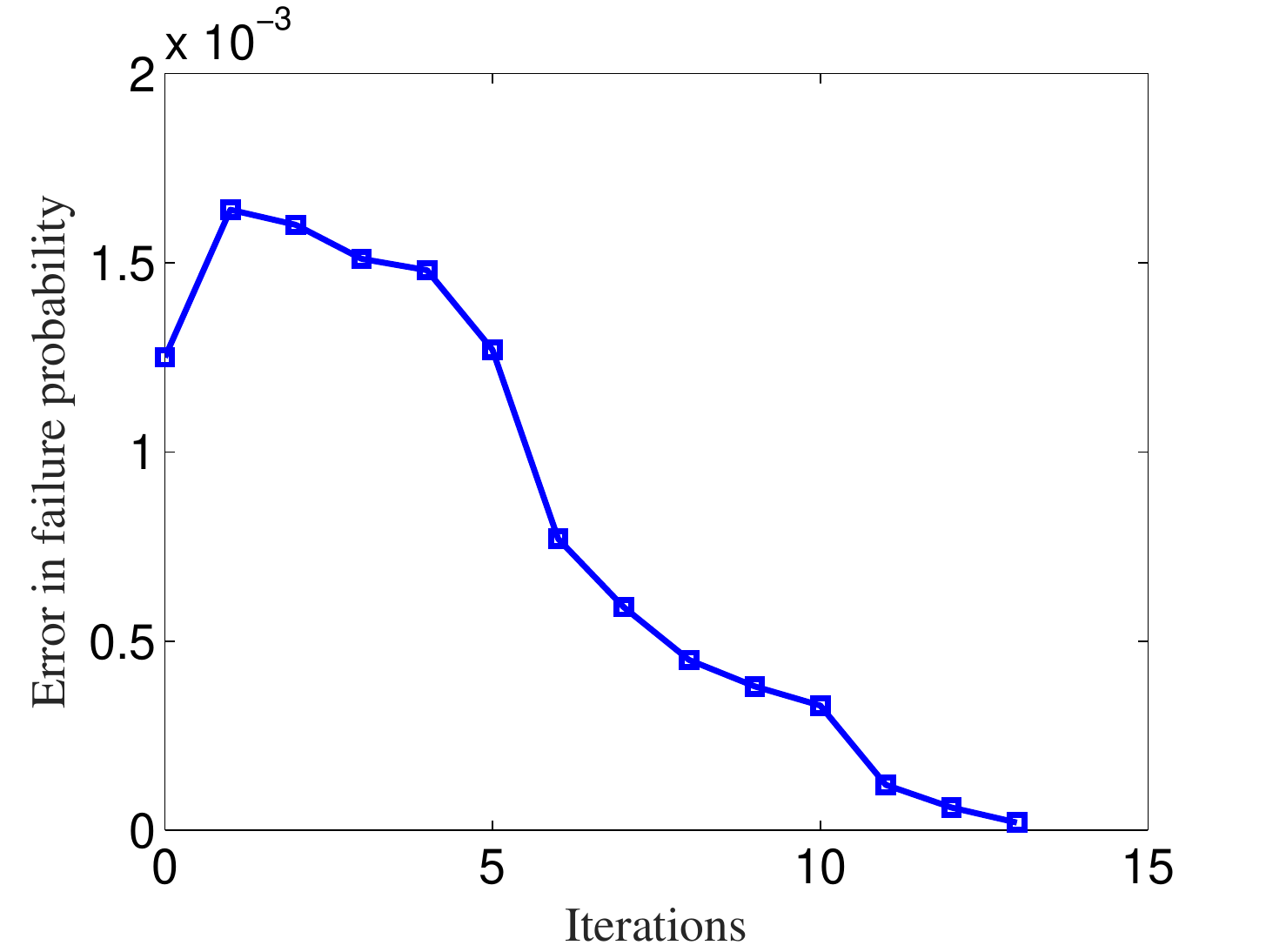}}
\caption{Left: the probability of detection error plotted against the number of design points.
Right: the error in the failure probability estimates plotted against the number of design points.}\label{f:err_4b}
\end{figure}

\begin{figure}
\centerline{\includegraphics[width=.45\textwidth]{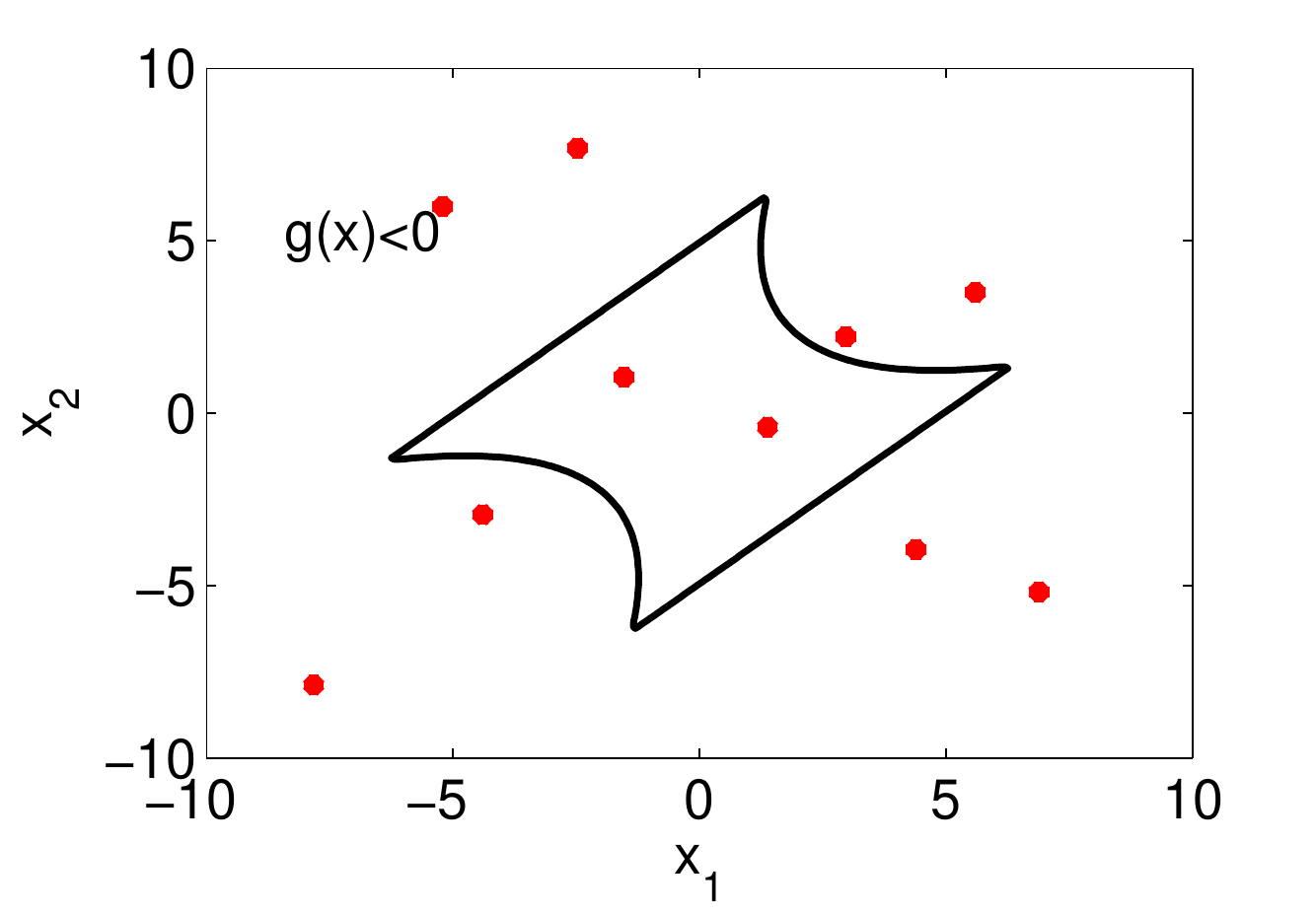}
\includegraphics[width=.45\textwidth]{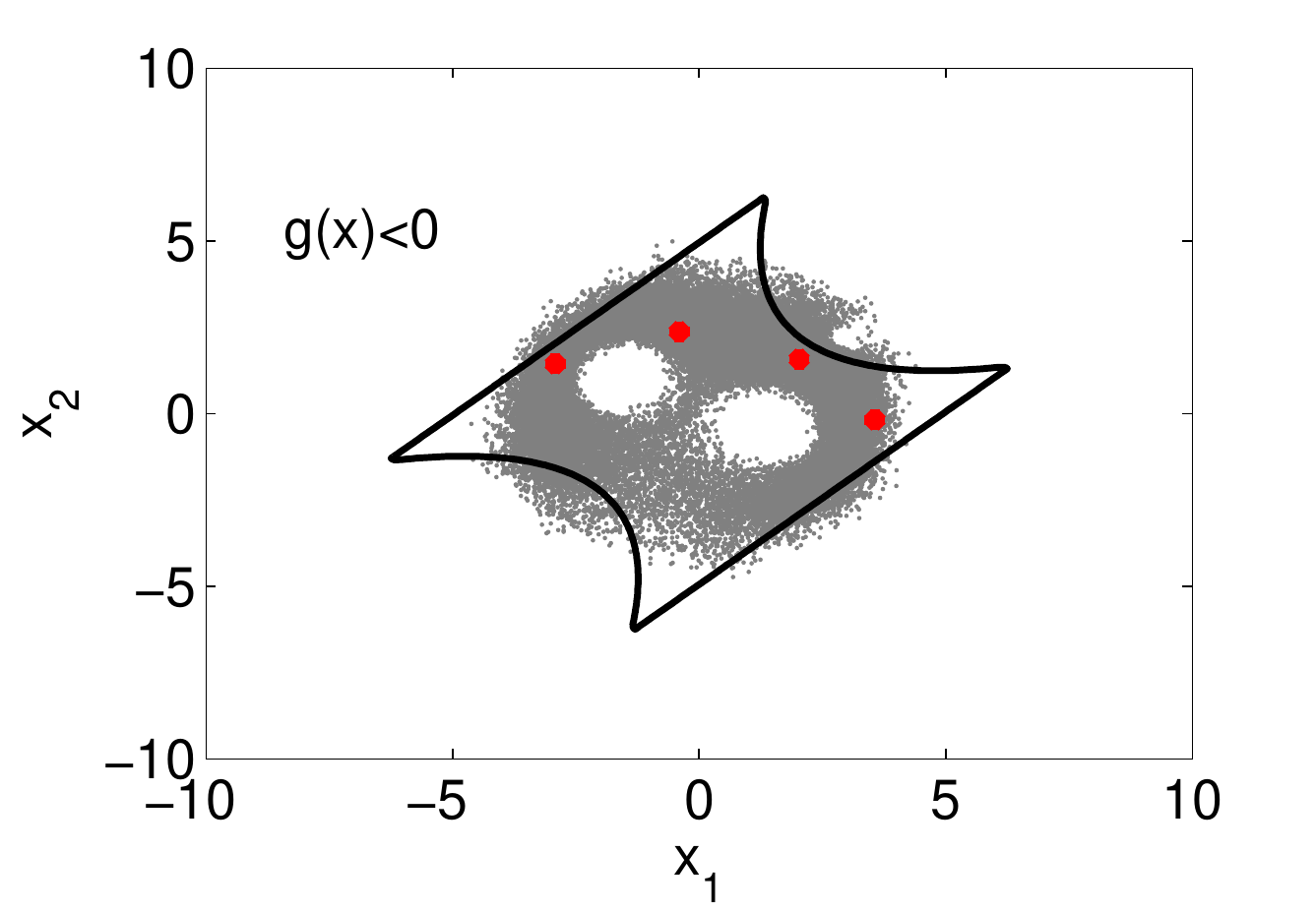}}
\centerline{\includegraphics[width=.45\textwidth]{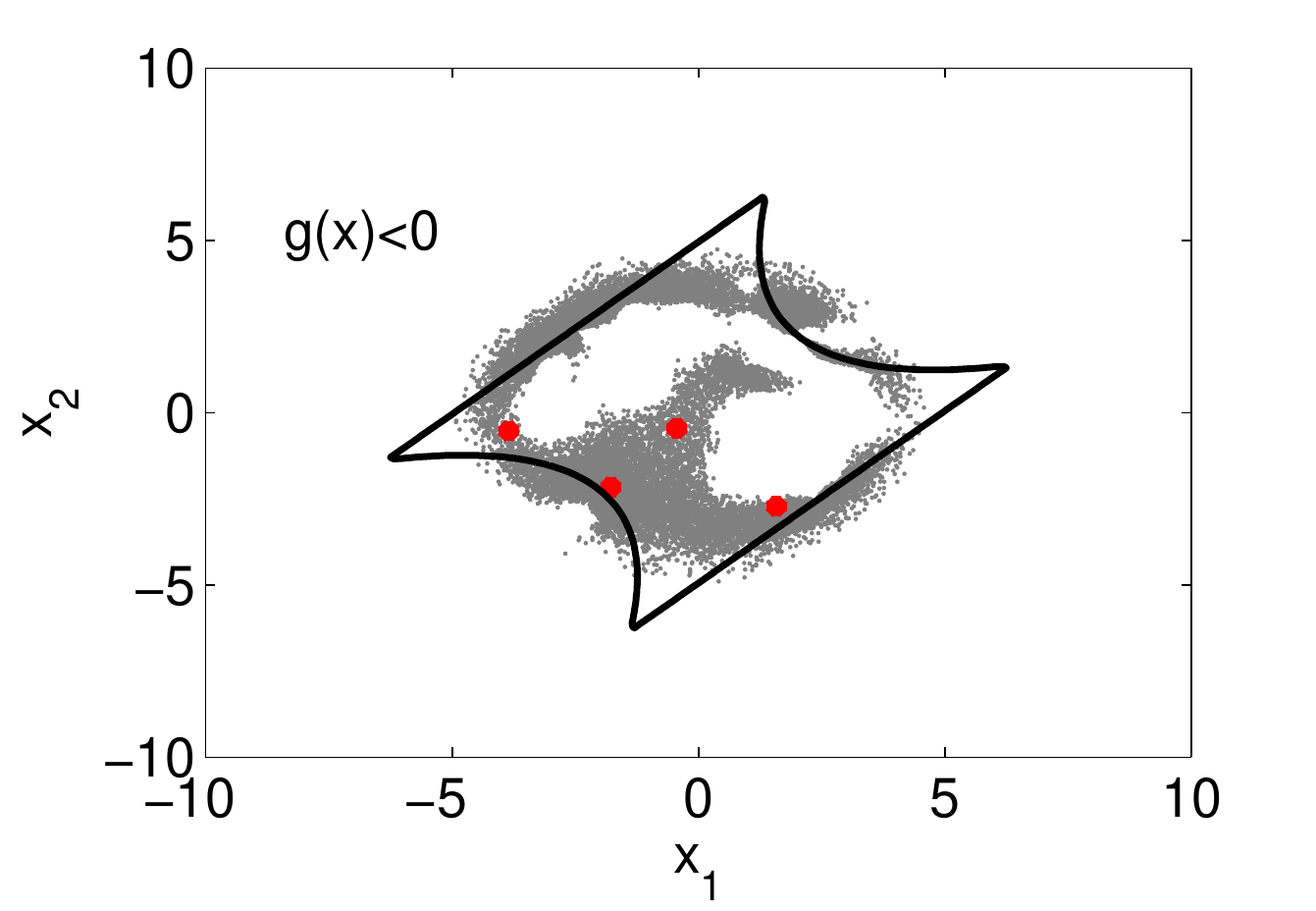}
\includegraphics[width=.45\textwidth]{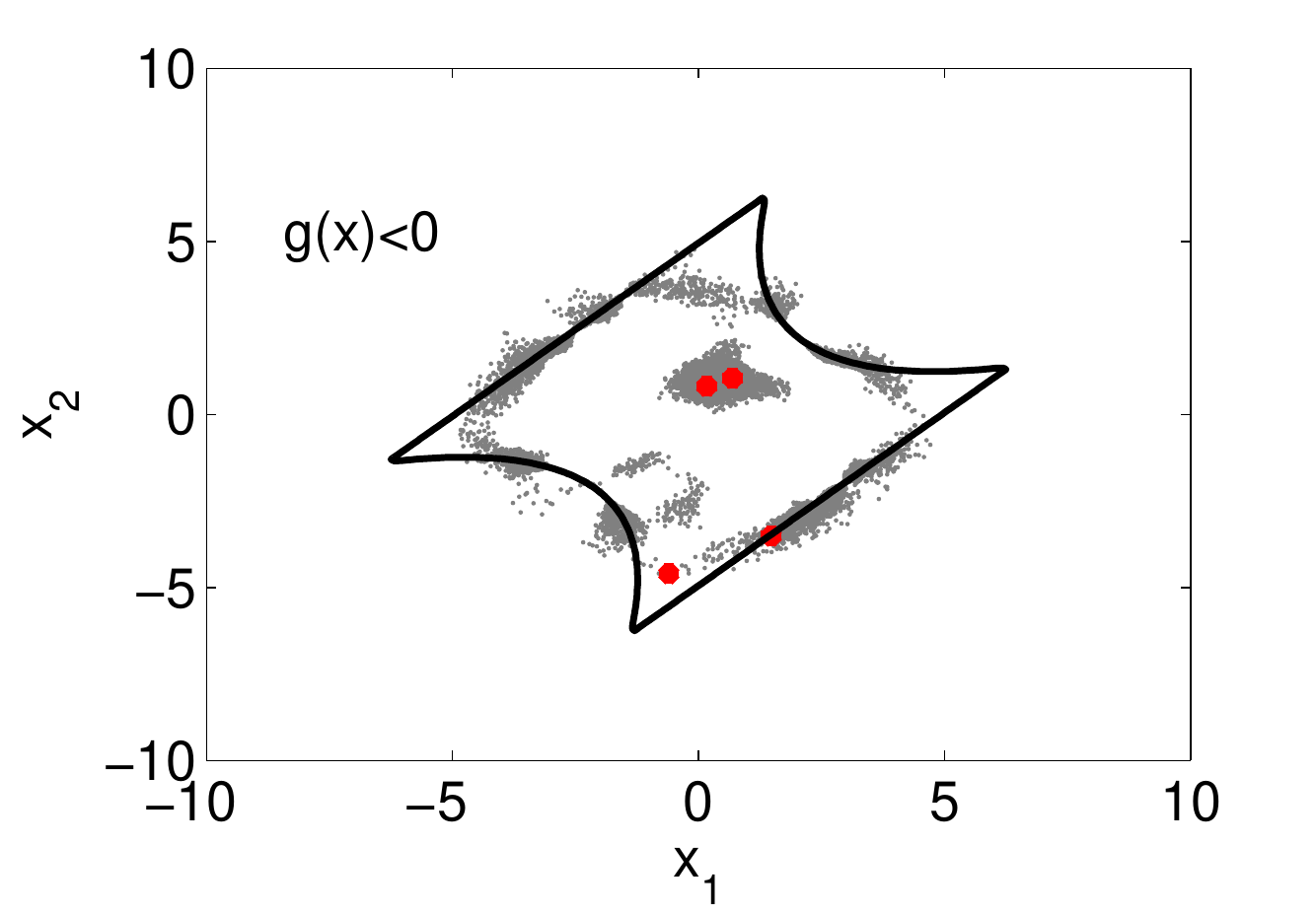}} 
\caption{ Top: the left figure shows the initial design points and the right one shows the design points determined in the first iteration.
Bottom: the left and the right figures are the design points determined in the 7th and the 13th iteration respectively. }\label{f:design}
\end{figure}

\begin{figure}
\centerline{\includegraphics[width=.47\textwidth]{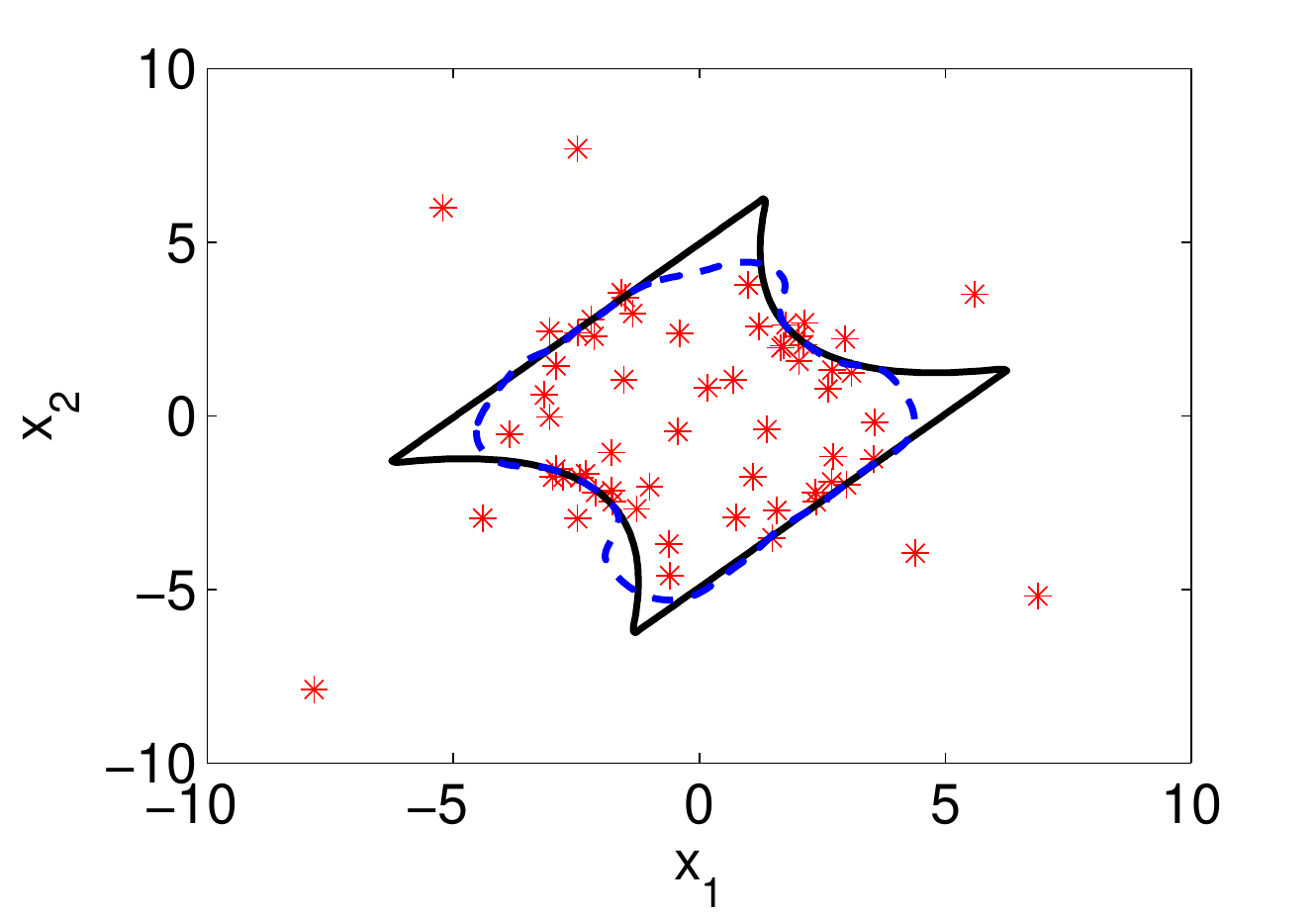}
\includegraphics[width=.47\textwidth]{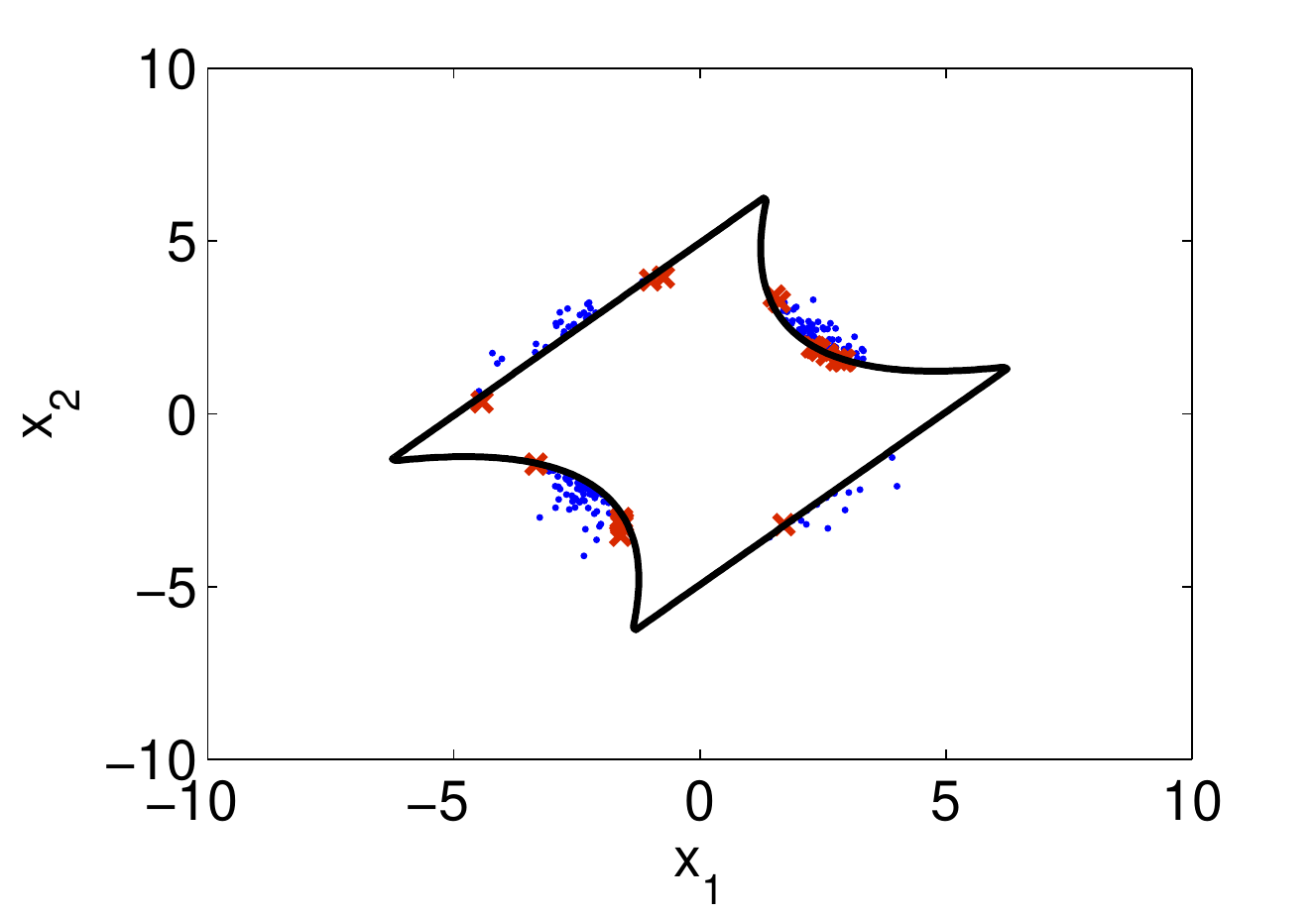}
} 
\caption{Left: all the design points (asterisks), the limit stage (solid line) and the GP surrogate (dashed line). 
Right: the dots are the points that are corrected identified as failures and the crosses are the mis-identified samples.
}\label{f:results}
\end{figure}

We use the squared exponential covariance function~\eqref{e:secov} in this example, but
with $a=0.2$ and $\beta=0.8$.
The prior mean of the GP surrogate is computed with a quadratic regression.
We perform the sequential design described in Section~\ref{s:method} with $4$ sampling points determined in each iteration.
The algorithm terminates in $13$ iterations, resulting in totally $62$ design points.
We plot the errors in failure probability estimation as a function of the number of iterations in Fig.~\ref{f:err_4b} (left),
and the mis-detection probability, also as a function of the number of iterations in Fig.~\ref{f:err_4b} (right).
In Figs.~\ref{f:design}, we plot the initial design points and the design points found in the first, the 7th, and the last iterations.
In Fig.~\ref{f:results} (left), we show all the 62 design points including both the initial points
and those found by our method. Also shown in the figure is the surrogate constructed with the obtained design points.
We can see that our method allocates more points near the boundary of the failure region than random sampling points.  
In Fig.~\ref{f:results} (right), we show the samples
that are incorrectly identified (crosses) and those are correctly identified as failures (dots).

We note that, the failure probability estimate obtained in the final iteration is $2.31\times10^{-3}$, while
the estimate of standard MC is $2.34\times10^{-3}$.
Also, compared to the MC simulation with true model, we find that totally $19$ samples are incorrectly classified: $11$ safe samples are identified as failures,
and $8$ failure samples are identified as safe ones.
The numerical results indicate that our method can obtain reliable estimates of the failure probability with a rather small number of sampling points.

\subsection{Clamped beam dynamics}
\begin{figure}
\centerline{\includegraphics[width=.7\textwidth]{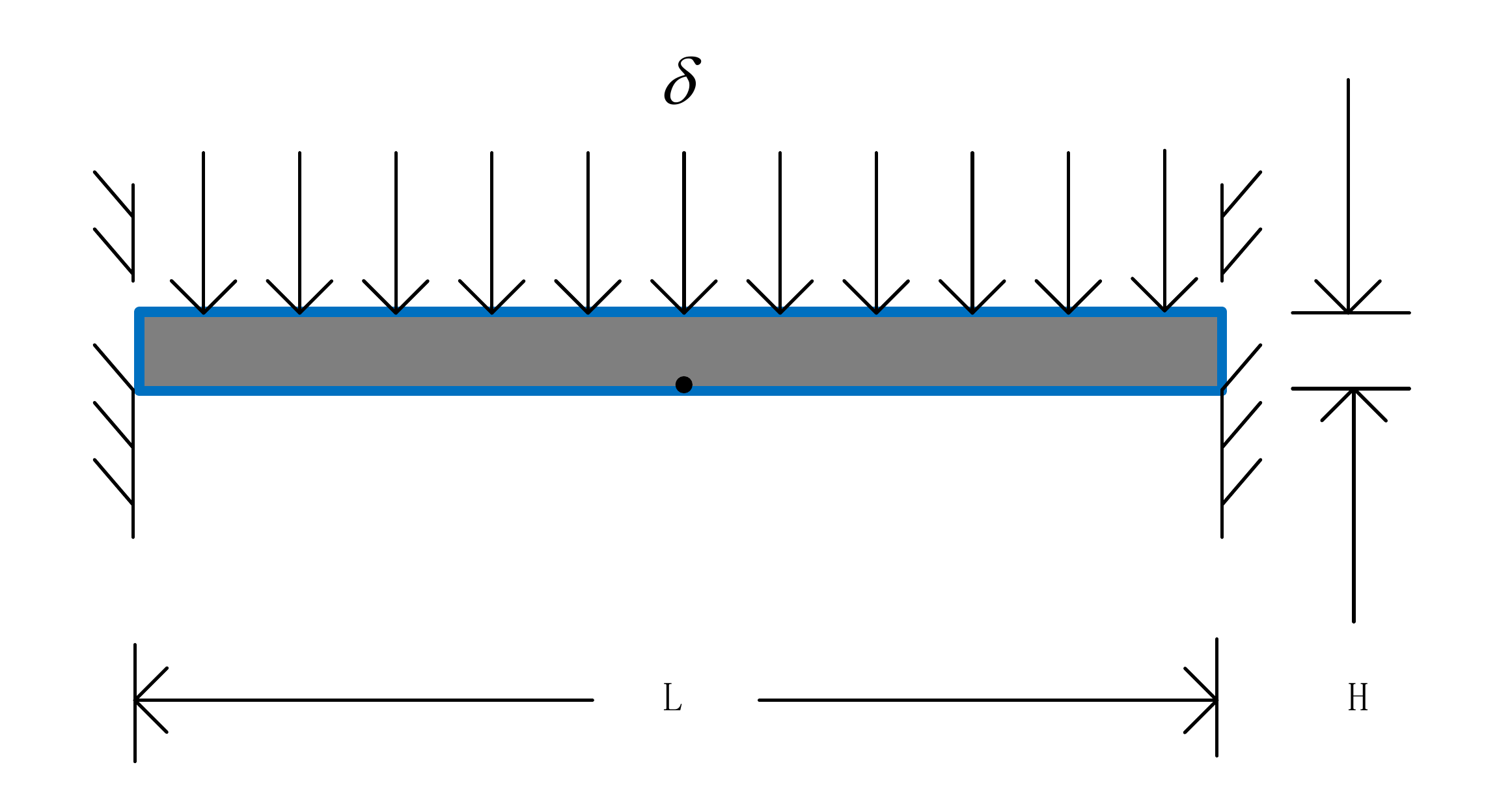}} 
\caption{Schematic illustration of the clamped beam.}\label{f:beam}
\end{figure}

Our last example is a practical problem which concerns the dynamic behavior of a beam clamped at both ends and a uniform pressure load is suddenly applied  {as shown in Fig.~\ref{f:beam}}.
We are interested in the dynamics of the deflections at the center of the beam caused by the pressure.

 The Lagrangian formulation of the equilibrium equations without body forces can be written
 $$ \rho \frac{d^2\-u}{dt^2} -\nabla \cdot (\-F \cdot \-S) = 0, $$
 where $\rho$ is the material density, $\-u$ is the displacement vector,
 $\-F$ is the deformation gradient,
\[ \-F=\frac{\partial u_i}{\partial x_k}
 + \delta_{ik}\]
 and $\-S$ is the second Piola-Kirchoff
 stress tensor.
%
%
 For simplicity, we assume  linear elastic constitutive
 relations and isotropic material.
As a result, the constitutive relations may be written in matrix
 form:
 $$
\-S= \left\{
  \begin{array}{c}
  S_{11} \\ S_{22} \\ S_{12}
  \end{array} \right\} =
  \left[
  \begin{array}{ccc}
  C_{11} &   C_{12} &   \\
  C_{12} &   C_{22} &   \\
  &   & 2 G_{12}  \\
  \end{array}  \right]
  \left\{
  \begin{array}{c}
  E_{11} \\  E_{22} \\ E_{12}
  \end{array} \right\}
 $$
 where 
 $$
 E_{ij} = \frac{1}{2} \left(
 \frac{\partial u_i}{\partial x_j} + \frac{\partial u_j}{\partial x_i} +
 \frac{\partial u_k}{\partial x_i} \frac{\partial u_k}{\partial x_j}
 \right),
 $$
and
 $$ C_{11} = C_{22} = \frac{E}{1 - \nu^2},\quad C_{12} =  \frac{E\nu}{1 - \nu^2},\quad G_{12} =  \frac{E}{2(1 + \nu)}.  $$
 Here $E$ is Young's modulus and $\nu$ is Poisson's ratio.
The initial conditions are
\[ u(0,\-x)=0,\quad \frac{\partial u}{\partial t}(0,\-x) = 0.\]
%
 Readers who are interested in more details about the Lagrangian
 formulation for nonlinear elasticity can consult, for example, \cite{malvern1969introduction}.
\begin{table}
\centering
\begin{tabular}{lccccccc}
\hline
parameter  & L & h & $\rho$ &E&$\nu$&$\delta$\\
\hline
mean&5&0.1&$7.77\time10^{-4}$&$3\times10^7$&0.3&200\\
\hline
variance& $6.25\times10^{-2}$&   $2.5\times10^{-5}$ & $1.51\times10^{-9}$ &   $3.6\times10^{13}$&   $2.25\times10^{-4}$&40\\
\hline
\end{tabular}
\medskip

{\caption{The mean and variance of the random parameters in the clamped beam problem.}}
\label{ta:beam}
\end{table}

\begin{figure}
\centerline{\includegraphics[width=.7\textwidth]{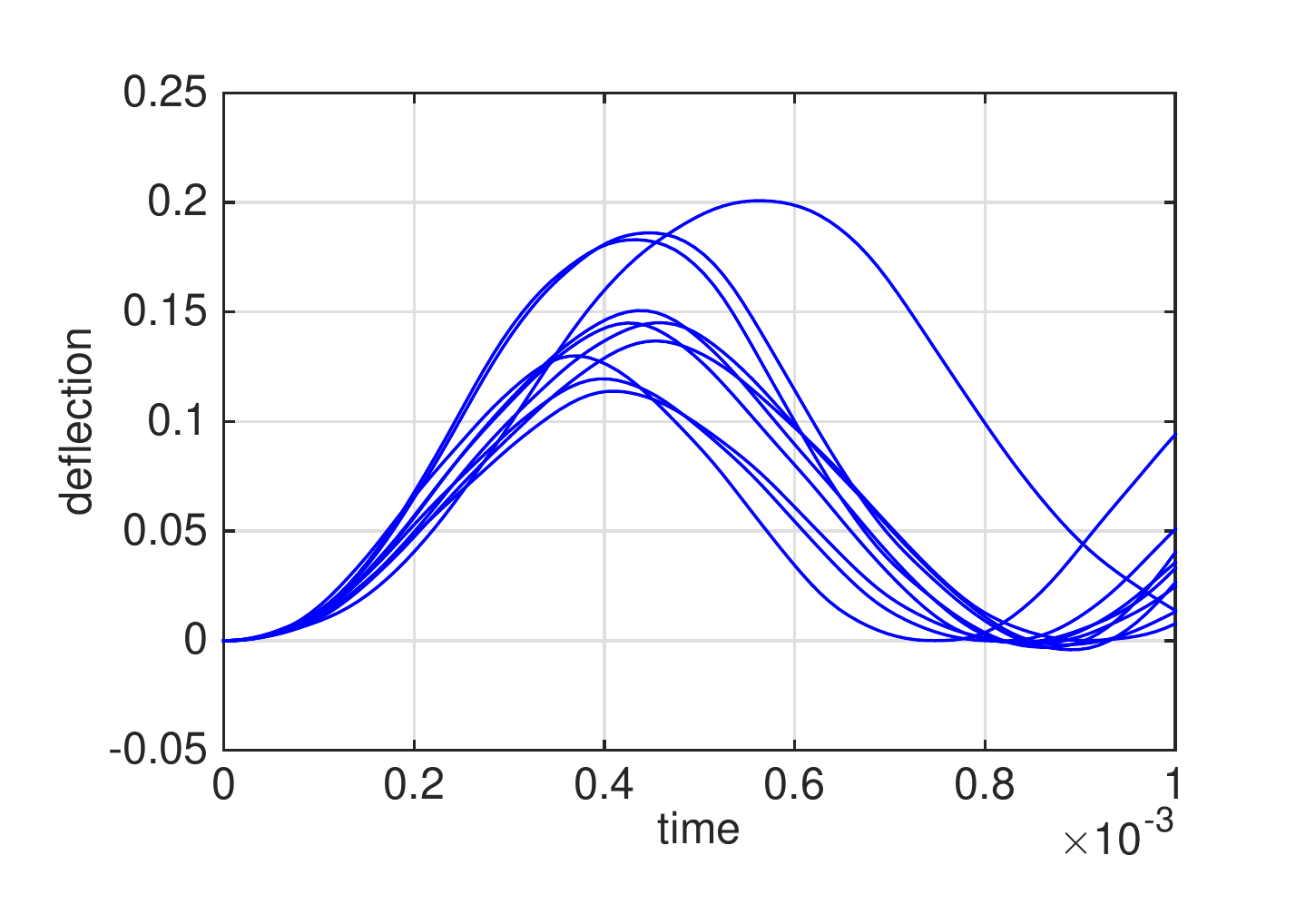}} 
\caption{Dynamics of the deflection at the beam center for ten randomly generated samples.}\label{f:deflection10}
\end{figure}

We assume in this examples that
the beam length $L$, the height $h$, the material density $\rho$,
the Young's module $E$,  the Poisson ratio $\nu$, and the applied load $\delta$ are random.
All the random parameters
follow a normal distribution and are independent to each other.
The means and the variances of the parameters are summarized in Table~1.
To demonstrate the statistical variations of the beam dynamics, we randomly generate 10 parameter samples 
and plot the 10 resulting beam dynamics in Fig.~\ref{f:deflection10}.  
We define the failure as the maximum deflection at the center of the beam being larger than a threshold value $u_{\max}$.
To be specific, we take $u_{\max} =0.23$.
We first run standard MC with $10^5$ samples to compute the failure probability, resulting in an estimate of $3.35\times10^{-3}$.

\begin{figure}
\centerline{\includegraphics[width=.45\textwidth]{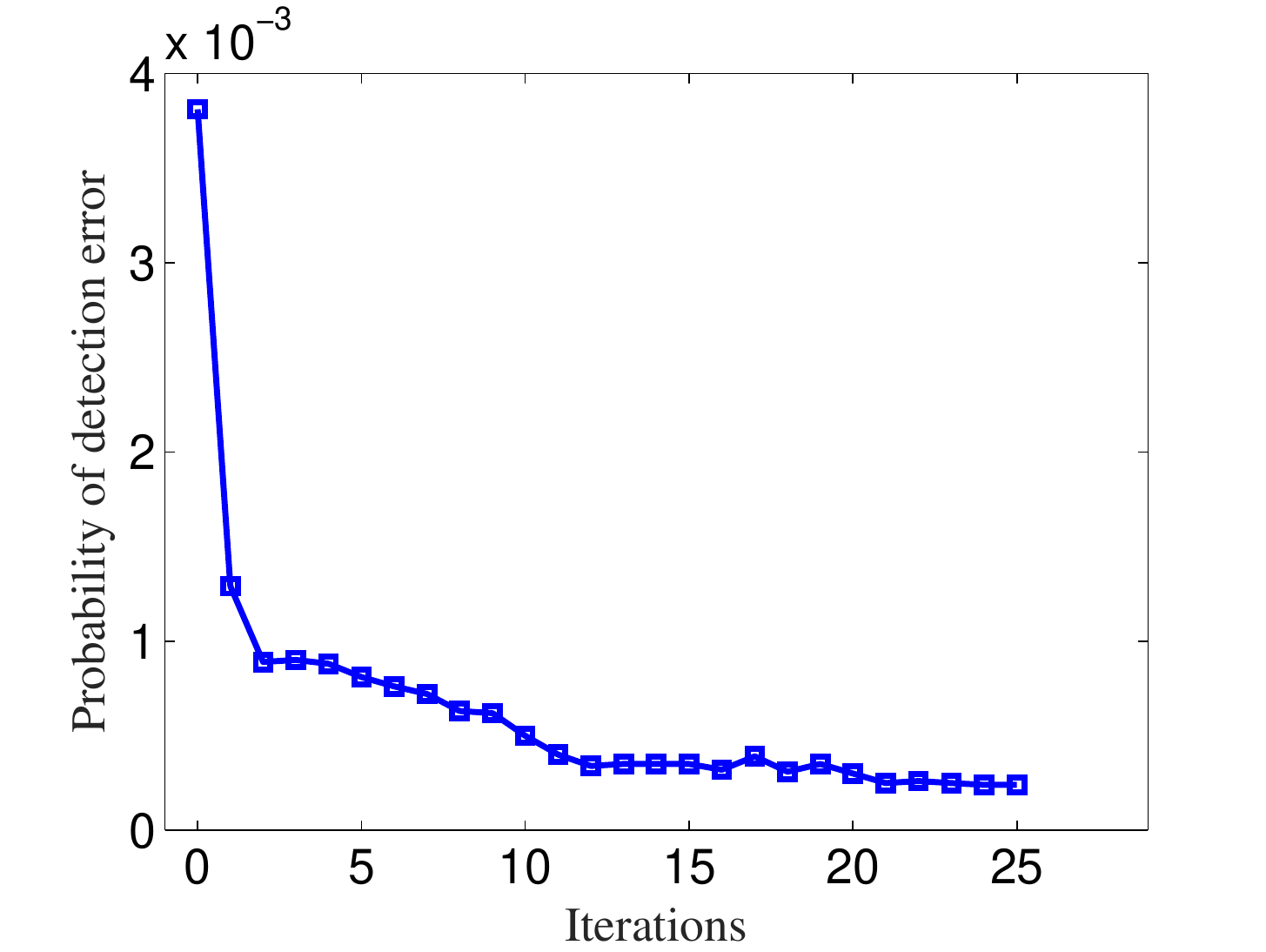}\includegraphics[width=.45\textwidth]{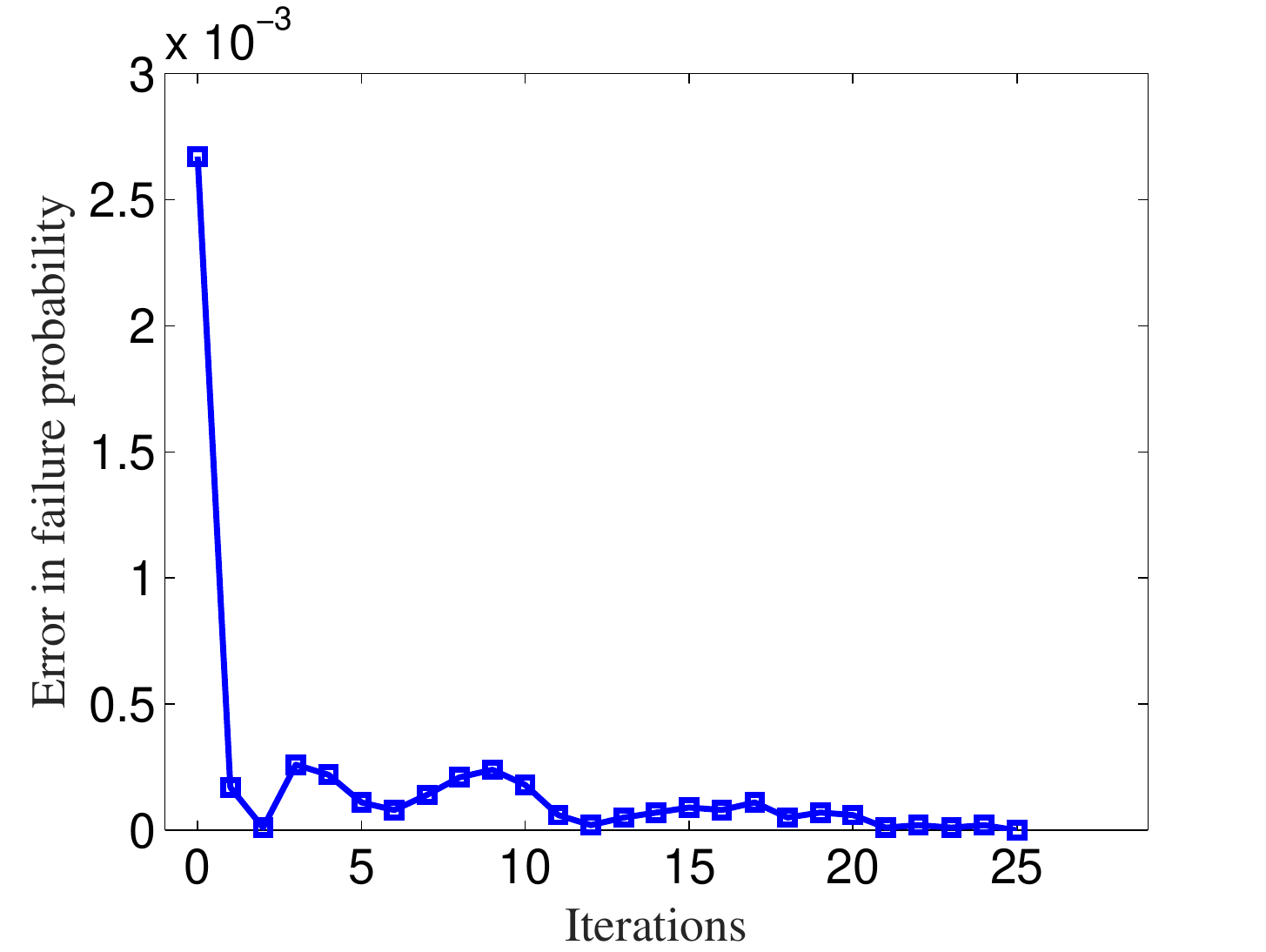}} 
\caption{Left: the probability of detection error plotted against the number of design points.
Right: the error in the failure probability estimates plotted against the number of design points.}\label{f:error_beam}
\end{figure}

In the GP method, we use $64$ initial sampling points drawn by the hyper Latin cube approach.
Our algorithm determines $4$ points in each iteration and it is terminated after 25 iterations, resulting in totally 164 sampling points.
As before, we plot the errors in failure probability estimation and the mis-detection probability, both against the number of iterations,
in Figs.~\ref{f:error_beam}.
In both plots we can see that the accuracy of the estimations improve rapidly as the number of sampling points increases, indicating 
that our methods can effectively identify good sampling points for this practical problem.
Our failure probability estimate is $3.35\times10^{-3}$ while $12$ points in the failure region is mis-identified as safe ones and
10 safe ones are mis-identified as failures.

\section{Conclusions}\label{s:conclusion}

In conclusion, we have presented a  {GP-based failure-detection} method to construct GP surrogate for failure probability estimations.
In particular, the method recasts the failure detection as inferring a contour $g(\-x)=0$ with Bayesian methods,
and then determines the optimal sampling locations  for the inference problem.
An efficient numerical implementation of the resulting design problem, based on
the SPSA algorithm, is also presented. 
With numerical examples, we demonstrate that the proposed LSI method can effectively and efficiently 
determine sampling points for failure detection and failure probability estimation. 

There are some improvements and extensions of the method that we plan to investigate in the future.
First, just like any  {GP-based} method, one must choose the  covariance kernels, the prior mean functions, and various hyperparameters.
We hope to develop effective methods to determine those functions and parameters specially for the failure probability estimation problems.
Secondly, as we have mentioned, the method shares a lot of common features of the SUR approach;
comprehensive comparison of the performance, and  detailed analysis of the advantages and limitations of the two methods, are
a very interesting problem to us.
Other possible improvements include to replace the greedy method with a dynamical programming approach in the sequential design,
and to use surrogates that combine both GP and polynomial chaos.
Finally, we note that, in addition to failure probability estimations, the LSI method can be applied to many
other problems as well. In particular, we plan to use the method to approximate the feasible regions in
constrained optimization problems~\cite{Picheny10adaptivedesigns}.

\section*{Acknowledgment}
The work was partially supported by the National Natural Science Foundation of China under grant number 11301337.  {G. Lin would like to acknowledge the support of the {U.S.} Department of Energy, Office of Science, Office of Advanced Scientific Computing Research, Applied Mathematics program as  part of the Multifaceted Mathematics for Complex Energy Systems (M$^2$ACS) project and part of the Collaboratory on Mathematics for Mesoscopic Modeling of Materials project, and NSF Grant DMS-1115887.}

\appendix
\section{The SPSA method}
 Here we briefly introduce the SPSA method,
in the context of our specific applications.
In each step, the method only uses two random
perturbations to estimate the gradient regardless of the problem's
dimension, which makes it particularly attractive for high dimensional problems.
Specifically, in step $j$, one first draw a $n_d$ dimensional random vector $\bm\Delta_j = [\delta_{j,1},...\delta_{j,n_d}]$, where $n_d = \mathrm{dim}(\-D)$,
from a prescribed distribution that is symmetric and of finite inverse moments.
The algorithm then updates the solution using the following equations:
\begin{subequations}
\begin{gather}
  \-D_{j+1}=\-D_j-a_j b_j(\-D_j), \\
  b_j(\-D_j)=\frac{\hat{U}(\-D_j+c_j\bm\Delta_j)-\hat{U}(\-D_j-c_j\bm\Delta_j)}{2c_j} \bm\Delta_j^{-1},
\end{gather}
\end{subequations}
where $\hat{U}(\cdot)$ is computed with Eq.~\eqref{e:uhat} and
 \[ \bm\Delta_j^{-1} = \left[
  \Delta^{-1}_{j,1},
    \ldots,
    \Delta^{-1}_{j,n_d}\right]^T, \quad
  a_k=\frac{a}{(A+k+1)^\alpha}, \quad c_k=\frac{c}{(k+1)^\gamma},
\]
with $A, \alpha, c$, and $\gamma$ being algorithm parameters.
Following the recommendation of [cite],  we choose $\bm\Delta_j \sim$ Bernoulli(0.5) and
$A=100,\,\alpha =0.602,\, c=1$, and $\gamma =0.101$ in this work.
\section{Derivation of Equation~(3.6)}

From Eq.~\eqref{e:Ud}, we have
 \begin{align}
  U(\-D) &=\int \int p(\-z| {\-y},\-D)\ln \left[\frac {p(\-z| {\-y},\-D)}{p(\-z)} \right]d\-z  \,p(\-y |\-D) \,d\-y \nonumber
 \\ &= \int \int
  \ln\left[\frac{p(\-y|\-z,\-D)}{p(\-y|\-D)}
  \right]p(\-y|\-z,\-D) \,
  p(\-z)\,d\-z \,d\-y  \nonumber
	 \\
	 & = \int \int
  \ln (p(\-y|\-z,\-D))
   p(\-y|\-z,\-D) \,
  p(\-z)\,d\-z \,d\-y \nonumber \\
  &- \int \int
  \ln[p(\-y|\-D)] p(\-y|\-z,\-D) \,
  p(\-z)\,d\-z \,d\-y
	\label{e:Ud2}
\end{align}

 Recall that $p(\-y|\-z,\-D)$ is multivariate normal distribution:
 \[p(\-y|\-{z},\-D)=\frac{1}{(2\pi)^{n/2}\cdot |\text{\sf C}|^{1/2}}\cdot
 \exp[-\frac{1}{2}(\-y-\-u)'\text{\sf C}^{-1}(\-y-\-u)],
\]
Because $\-u$ and $\text{\sf C}$ only depend on ${\-z} $ and $\-D$ (
$\mu=\mu({z},d),\text{\sf C}=\text{\sf C}({z},d)$ ),
  we change of variable:
  \[\-s=\text{\sf C}^{-1/2}(\-y-\-u)\]
\begin{flalign}
\int \int
 & \ln (p(\-y|\-z,\-D))
   p(\-z|\-z,\-D) \,
  p(\-z)\,d\-z \,d\-y \nonumber \\
&= \int\int \left( -\frac{n}2\ln{(2\pi) -\frac12\ln|\text{\sf C}|}
 -\frac{1}{2}(\-y-\-u)'\text{\sf C}^{-1}(\-y-\-u)\right)\nonumber \\
 &\qquad \times\frac{1}{(2\pi)^{n/2} |\text{\sf C}|^{1/2}}
 \exp[-\frac{1}{2}(\-y-\-u)'\text{\sf C}^{-1}(\-y-\-u)]
 p({\-z})d\-z d\-y\nonumber \\
&=\int\int \left( -\frac{n}2\ln{(2\pi) -\frac12\ln|\text{\sf C}|}
 -\frac{1}{2}\-s'\-s\right)\frac{1}{(2\pi)^{n/2}}
 \exp(-\frac{1}{2}\-s'\-s)
 p({\-z}) d{\-z} d\-s  \nonumber\\
&=-\frac{n}2\ln(2\pi)-\frac12\int \ln|\text{\sf C}|d\-z -\int\int \left(
 \frac{1}{2}\-s'\-s\right)\frac{1}{(2\pi)^{n/2}}
 \exp(-\frac{1}{2}\-s'\-s)
 p({\-z}) d{\-z} d\-s \nonumber\\
&=-\frac12\int \ln|\text{\sf C}| p(\-z)d\-z -\int \left(
 \frac{1}{2}\-s'\-s\right)\frac{1}{(2\pi)^{n/2}}
 \exp(-\frac{1}{2}\-s'\-s)
  d\-s -\frac{n}2\ln(2\pi) \label{e:ud2}
  \end{flalign}
  Note that the second integral on the right hand side of Eq.~\eqref{e:ud2} actually does
not depend on $\-D$ and so we can define a constant $Z$ such that
\[Z=-\int\int \left(
 \frac{1}{2}\-s'\-s\right)\frac{1}{(2\pi)^{n/2}}
 \exp(-\frac{1}{2}\-s'\-s)
  d\-s -\frac{n}2\ln(2\pi) .\]
It follows immediately that
\[\int \int \ln (p(\-y|\-z,\-D))
   p(\-z|\-z,\-D) \,
  p(\-z)\,d\-z \,d\-y = -\frac12\E_z[\ln|\-C|] +Z,
  \]
which in turns yields Eq.~\eqref{e:Ud3}.

\bibliographystyle{plain}
\bibliography{reliability}

\begin{thebibliography}{10}

\bibitem{ahmad1976nonparametric}
Ibrahim Ahmad, Pi-Erh Lin, et~al.
\newblock A nonparametric estimation of the entropy for absolutely continuous
  distributions (corresp.).
\newblock {\em Information Theory, IEEE Transactions on}, 22(3):372--375, 1976.

\bibitem{AuB99}
S.K. Au and J.~Beck.
\newblock A new adaptive importance sampling scheme for reliability
  calculations.
\newblock {\em Struct. Safety}, 21(2):135--158, 1999.

\bibitem{AuB01}
S.K. Au and J.~Beck.
\newblock Estimation of small failure probabilities in high dimensions by
  subset simulation.
\newblock {\em Prob. Eng. Mech.}, 16:263--277, 2001.

\bibitem{balesdent2013kriging}
Mathieu Balesdent, Jerome Morio, and Julien Marzat.
\newblock Kriging-based adaptive importance sampling algorithms for rare event
  estimation.
\newblock {\em Structural Safety}, 44:1--10, 2013.

\bibitem{bect2012sequential}
Julien Bect, David Ginsbourger, Ling Li, Victor Picheny, and Emmanuel Vazquez.
\newblock Sequential design of computer experiments for the estimation of a
  probability of failure.
\newblock {\em Statistics and Computing}, 22(3):773--793, 2012.

\bibitem{beirlant1997nonparametric}
Jan Beirlant, Edward~J Dudewicz, L{a}szl{o} Gyorfi, and Edward~C Van~der
  Meulen.
\newblock Nonparametric entropy estimation: An overview.
\newblock {\em International Journal of Mathematical and Statistical Sciences},
  6(1):17--39, 1997.

\bibitem{bichon2008efficient}
Barron~J Bichon, Michael~S Eldred, Laura~Painton Swiler, Sandaran Mahadevan,
  and John~M McFarland.
\newblock Efficient global reliability analysis for nonlinear implicit
  performance functions.
\newblock {\em AIAA journal}, 46(10):2459--2468, 2008.

\bibitem{bilionis2012multi}
Ilias Bilionis and Nicholas Zabaras.
\newblock Multi-output local gaussian process regression: Applications to
  uncertainty quantification.
\newblock {\em Journal of Computational Physics}, 231(17):5718--5746, 2012.

\bibitem{bilionis2013multi}
Ilias Bilionis, Nicholas Zabaras, Bledar~A Konomi, and Guang Lin.
\newblock Multi-output separable gaussian process: Towards an efficient, fully
  bayesian paradigm for uncertainty quantification.
\newblock {\em Journal of Computational Physics}, 241:212--239, 2013.

\bibitem{BucherB90}
C.G. Bucher and U.~Bourgund.
\newblock A fast and efficient response surface approach for structural
  reliability problems.
\newblock {\em Struct. Safety}, 7:57--66, 1990.

\bibitem{cerou2012sequential}
Fr{e}d{e}ric C{e}rou, Pierre Del~Moral, Teddy Furon, and Arnaud Guyader.
\newblock Sequential monte carlo for rare event estimation.
\newblock {\em Statistics and Computing}, 22(3):795--808, 2012.

\bibitem{chevalier2014fast}
Clement Chevalier, David Ginsbourger, Julien Bect, Emmanuel Vazquez, Victor
  Picheny, and Yann Richet.
\newblock Fast parallel kriging-based stepwise uncertainty reduction with
  application to the identification of an excursion set.
\newblock {\em Technometrics}, 56(4), 2014.

\bibitem{chevalier2014corrected}
Cl{\'e}ment Chevalier, David Ginsbourger, and Xavier Emery.
\newblock Corrected kriging update formulae for batch-sequential data
  assimilation.
\newblock In {\em Mathematics of Planet Earth}, pages 119--122. Springer, 2014.

\bibitem{Chevalier20141021}
Cl\'ement Chevalier, Victor Picheny, and David Ginsbourger.
\newblock Kriginv: An efficient and user-friendly implementation of
  batch-sequential inversion strategies based on kriging.
\newblock {\em Computational Statistics and Data Analysis}, 71:1021 -- 1034,
  2014.

\bibitem{ditlevsen1996structural}
Ove Ditlevsen and Henrik~O Madsen.
\newblock {\em Structural reliability methods}, volume 178.
\newblock Wiley New York, 1996.

\bibitem{dubourg2013metamodel}
Vincent Dubourg, B~Sudret, and F~Deheeger.
\newblock Metamodel-based importance sampling for structural reliability
  analysis.
\newblock {\em Probabilistic Engineering Mechanics}, 33:47--57, 2013.

\bibitem{echard2011ak}
B~Echard, N~Gayton, and M~Lemaire.
\newblock Ak-mcs: an active learning reliability method combining kriging and
  monte carlo simulation.
\newblock {\em Structural Safety}, 33(2):145--154, 2011.

\bibitem{emery2009kriging}
Xavier Emery.
\newblock The kriging update equations and their application to the selection
  of neighboring data.
\newblock {\em Computational Geosciences}, 13(3):269--280, 2009.

\bibitem{Engelund93}
S.~Engelund and R.~Rackwitz.
\newblock A benchmark study on importance sampling techniques in structural
  reliability.
\newblock {\em Struct. Safety}, 12:255--276, 1993.

\bibitem{Faravelli89}
L.~Faravelli.
\newblock Response surface approach for reliability analysis.
\newblock {\em J. Eng. Mech.}, 115(12):2763--2781, 1989.

\bibitem{GaytonBL03}
N.~Gayton, J.M. Bourinet, and M.~Lemaire.
\newblock {CQ2RS}: a new statistical approach to the response surface method
  for reliability analysis.
\newblock {\em Struct. Safety}, 25:99--121, 2003.

\bibitem{GuptaM04}
S.~Gupta and C.S. Manohar.
\newblock An improved response surface method for the determination of failure
  probability and importance measures.
\newblock {\em Struct. Safety}, 26:123--139, 2004.

\bibitem{hall1993estimation}
Peter Hall and Sally~C Morton.
\newblock On the estimation of entropy.
\newblock {\em Annals of the Institute of Statistical Mathematics},
  45(1):69--88, 1993.

\bibitem{Huan2013}
Xun Huan and Youssef~M. Marzouk.
\newblock {Simulation-based optimal Bayesian experimental design for nonlinear
  systems}.
\newblock {\em Journal of Computational Physics}, 232(1):288--317, 2013.

\bibitem{joe1989estimation}
Harry Joe.
\newblock Estimation of entropy and other functionals of a multivariate
  density.
\newblock {\em Annals of the Institute of Statistical Mathematics},
  41(4):683--697, 1989.

\bibitem{Krause2008}
Andreas Krause, Ajit Singh, and Carlos Guestrin.
\newblock {Near-Optimal Sensor Placements in Gaussian Processes-Theory,
  Efficient Algorithms and Empirical Studies}.
\newblock {\em The Journal of Machine Learning Research}, 9(May):235--284,
  2008.

\bibitem{li2011efficient}
Jing Li, Jinglai Li, and Dongbin Xiu.
\newblock An efficient surrogate-based method for computing rare failure
  probability.
\newblock {\em Journal of Computational Physics}, 230(24):8683--8697, 2011.

\bibitem{li2012bayesian}
Ling Li, Julien Bect, and Emmanuel Vazquez.
\newblock Bayesian subset simulation: a kriging-based subset simulation
  algorithm for the estimation of small probabilities of failure.
\newblock {\em arXiv preprint arXiv:1207.1963}, 2012.

\bibitem{long2013fast}
Quan Long, Marco Scavino, Ra{\'u}l Tempone, and Suojin Wang.
\newblock Fast estimation of expected information gains for bayesian
  experimental designs based on laplace approximations.
\newblock {\em Computer Methods in Applied Mechanics and Engineering},
  259:24--39, 2013.

\bibitem{malvern1969introduction}
Lawrence~E Malvern.
\newblock {\em Introduction to the Mechanics of a Continuous Medium}.
\newblock Number Monograph. 1969.

\bibitem{mclachlan2004finite}
Geoffrey McLachlan and David Peel.
\newblock {\em Finite mixture models}.
\newblock John Wiley \& Sons, 2004.

\bibitem{melchers1989importance}
RE~Melchers.
\newblock Importance sampling in structural systems.
\newblock {\em Structural safety}, 6(1):3--10, 1989.

\bibitem{OHagan1978}
Anthony O'Hagan and JFC Kingman.
\newblock Curve fitting and optimal design for prediction.
\newblock {\em Journal of the Royal Statistical Society. Series B
  (Methodological)}, pages 1--42, 1978.

\bibitem{oliver1990kriging}
Margaret~A Oliver and Richard Webster.
\newblock Kriging: a method of interpolation for geographical information
  systems.
\newblock {\em International Journal of Geographical Information System},
  4(3):313--332, 1990.

\bibitem{Picheny10adaptivedesigns}
V.~Picheny, D.~Ginsbourger, O.~Roustant, R.~T. Haftka, and N.~H. Kim.
\newblock Adaptive designs of experiments for accurate approximation of a
  target region.
\newblock {\em Journal of Mechanical Design}, 2010.

\bibitem{Pulch_2008}
R.~Pulch.
\newblock Polynomial chaos for the computation of failure probabilities in
  periodic problems.
\newblock In J.~Roos and L.~Costa, editors, {\em Scientific Computing in
  Electrical Engineering SCEE 2008}, 2010.

\bibitem{RajashekharE93}
M.R. Rajashekhar and B.R. Ellingwood.
\newblock A new look at the response surface approach for reliability analysis.
\newblock {\em Struct. Safety}, 12:205--220, 1993.

\bibitem{RubinsteinK_04}
R.Y. Rubinstein and D.P. Kroese.
\newblock {\em The cross-entropy method}.
\newblock Springer Science+Business Media, Inc., New York, NY, 2004.

\bibitem{SchuellerPK04}
G.I. Schueller, H.J. Pradlwarter, and P.S. Koutsourelakis.
\newblock A critical appraisal of reliability estimation procedures for high
  dimensions.
\newblock {\em Prob. Eng. Mech.}, 19:463--474, 2004.

\bibitem{simpson2001kriging}
Timothy~W Simpson, Timothy~M Mauery, John~J Korte, and Farrokh Mistree.
\newblock Kriging models for global approximation in simulation-based
  multidisciplinary design optimization.
\newblock {\em AIAA journal}, 39(12):2233--2241, 2001.

\bibitem{spall1992multivariate}
James~C Spall.
\newblock Multivariate stochastic approximation using a simultaneous
  perturbation gradient approximation.
\newblock {\em Automatic Control, IEEE Transactions on}, 37(3):332--341, 1992.

\bibitem{spall1998implementation}
James~C Spall.
\newblock Implementation of the simultaneous perturbation algorithm for
  stochastic optimization.
\newblock {\em Aerospace and Electronic Systems, IEEE Transactions on},
  34(3):817--823, 1998.

\bibitem{wang2015cross}
Hui Wang and Xiang Zhou.
\newblock A cross-entropy scheme for mixtures.
\newblock {\em ACM Transactions on Modeling and Computer Simulation (TOMACS)},
  25(1):6, 2015.

\bibitem{williams2006gaussian}
Christopher~KI Williams and Carl~Edward Rasmussen.
\newblock Gaussian processes for machine learning.
\newblock {\em the MIT Press}, 2(3):4, 2006.

\end{thebibliography}

\end{document}